%% file: optext.tex
\documentclass[11pt]{article}
\usepackage{amsmath,amsthm,bm, amscd}
\usepackage{fullpage}
\usepackage[nofullpage,full,titlepage,nousetoc,nouselot,nouselof,hylinks,final]{boaz}

\usepackage{color}

\usepackage{graphicx}

\newcommand{\supp}{\mathsf{supp}}

\newcommand{\iext}{\mathsf{IExt}}

\newcommand{\zo}{\bits}

\newcommand{\acb}{\mathsf{AdvCB}}
\newcommand{\adg}{\mathsf{AdvGen}}
\newcommand{\RS}{\mathsf{RS}}

\newcommand{\snmExt}{\textnormal{snmExt}}
\newcommand{\nipm}{\textnormal{NIPM}}
\newcommand{\A}{\mathcal{A}}
\newcommand{\scirc}{\hspace{0.1cm}\circ \hspace{0.1cm}}
\newcommand{\X}{\mathbf{X}}
\newcommand{\rr}{\mathbf{R}}
\newcommand{\Y}{\mathbf{Y}}
\newcommand{\U}{\mathbf{U}}
\newcommand{\snm}{\mathsf{NM}}
\newcommand{\flip}{\textnormal{flip-flop}}

\def\calX{{\mathcal X}}
\def\calY{{\mathcal Y}}

\DeclareMathOperator{\expect}{E}

\newcommand{\laext}{\mathsf{laExt}}
\newcommand{\samp}{\mathsf{Samp}}

\theoremstyle{definition}

\newcommand{\eps}{\epsilon}

\newcommand{\Cond}{\mathsf{Cond}}

\newcommand{\czext}{\mathsf{CZExt}}

\newcommand{\Supp}{\mathsf{Supp}}

\newcommand{\Ext}{\mathsf{Ext}}

\newcommand{\nmExt}{\mathsf{nmExt}}
\newcommand{\mac}{\mathsf{MAC}}
\newcommand{\bip}{\mathsf{IP}}
\newcommand{\Enc}{\mathsf{Enc}}
\newcommand{\Dec}{\mathsf{Dec}}

\newcommand{\TExt}{\mathsf{TExt}}
\newcommand{\zuc}{\Cond}
\newcommand{\scond}{\mathsf {Scond}}

\newcommand{\rip}{\mathsf{RIP}}

\newcommand{\BI}{\begin{itemize}}
\newcommand{\EI}{\end{itemize}}
\newcommand{\BE}{\begin{enumerate}}
\newcommand{\EE}{\end{enumerate}}

\newtheorem{thm}{Theorem}      
\newcommand{\BT}{\begin{theorem}}   \newcommand{\ET}{\end{theorem}}
\newcommand{\BD}{\begin{definition}}   \newcommand{\ED}{\end{definition}}
\newcommand{\BCR}{\begin{corollary}} \newcommand{\ECR}{\end{corollary}}
\newtheorem{constr}[thm]{Construction}
\newcommand{\BCT}{\begin{constr}} \newcommand{\ECT}{\end{constr}}
\newcommand{\BL}{\begin{lemma}}   \newcommand{\EL}{\end{lemma}}

\newcommand{\BP}{\begin{proposition}}   \newcommand{\EP}{\end{proposition}}
\newcommand{\BCM}{\begin{claim}}   \newcommand{\ECM}{\end{claim}}
\newcommand{\BF}{\begin{fact}}   \newcommand{\EF}{\end{fact}}
\newcommand{\BA}{\begin{assumption}}   \newcommand{\EA}{\end{assumption}}
\newcommand{\tabincell}[2]{\begin{tabular}{@{}#1@{}}#2\end{tabular}}

\def\eps{\varepsilon}

\def\le{\leqslant} \def\ge{\geqslant}

\makeatletter
\def\ExtendSymbol#1#2#3#4#5{\ext@arrow 0099{\arrowfill@#1#2#3}{#4}{#5}}
\def\RightExtendSymbol#1#2#3#4#5{\ext@arrow 0359{\arrowfill@#1#2#3}{#4}{#5}}
\def\LeftExtendSymbol#1#2#3#4#5{\ext@arrow 6095{\arrowfill@#1#2#3}{#4}{#5}}
\makeatother
\newcommand\llrightarrow[2][]{\RightExtendSymbol{-}{-}{\rightarrow}{#1}{#2}}

\newcommand\llleftarrow[2][]{\RightExtendSymbol{\leftarrow}{-}{-}{#1}{#2}}

\newcommand{\hinf}{H_\infty}
\newcommand{\thinf}{\widetilde{H}_\infty}

\begin{document}

\begin{titlepage}
\def\thepage{}

\date{}
\title{
Improved Non-Malleable Extractors, Non-Malleable Codes and Independent Source Extractors}

\author{
Xin Li \thanks{Supported in part by NSF Grant CCF-1617713.}\\
Department of Computer Science\\
Johns Hopkins University\\
Baltimore, MD 21218, U.S.A.\\
lixints@cs.jhu.edu
}

\maketitle \thispagestyle{empty}

\begin{abstract}
In this paper we give improved constructions of several central objects in the literature of randomness extraction and tamper-resilient cryptography. Our main results are:

(1) An explicit seeded non-malleable extractor with error $\e$ and seed length $d=O(\log n)+O(\log(1/\e)\log \log (1/\e))$, that supports min-entropy $k=\Omega(d)$ and outputs $\Omega(k)$ bits. Combined with the protocol in \cite{DW09}, this gives a two round privacy amplification protocol with optimal entropy loss in the presence of an active adversary, for all security parameters up to $\Omega(k/\log k)$, where $k$ is the min-entropy of the shared weak random source. Previously, the best known seeded non-malleable extractors require seed length and min-entropy $O(\log n)+\log(1/\e)2^{O{\sqrt{\log \log (1/\e)}}}$ \cite{CL16, Coh16}, and only give two round privacy amplification protocols with optimal entropy loss for security parameter up to $k/2^{O(\sqrt{\log k})}$.

(2) An explicit non-malleable two-source extractor for min-entropy $k \geq (1-\gamma)n$, some constant $\gamma>0$, that outputs $\Omega(k)$ bits with error $2^{-\Omega(n/\log n)}$. We further show that we can efficiently uniformly sample from the pre-image of any output of the extractor. Combined with the connection in \cite{CG14b} this gives a non-malleable code in the two-split-state model with relative rate $\Omega(1/\log n)$. This exponentially improves previous constructions, all of which only achieve rate $n^{-\Omega(1)}$.\footnote{The work of Aggarwal et.\ al \cite{ADKO15} had a construction which ``achieves" constant rate, but recently the author found an error in their proof.} 

(3) Combined with the techniques in \cite{BDT16}, our non-malleable extractors give a two-source extractor for min-entropy $O(\log n \log \log n)$, which also implies a $K$-Ramsey graph on $N$ vertices with $K=(\log N)^{O(\log \log \log N)}$. Previously the best known two-source extractor in \cite{BDT16} requires min-entropy $\log n 2^{O(\sqrt{\log n})}$, which gives a Ramsey graph with $K=(\log N)^{2^{O(\sqrt{\log \log \log N})}}$. We further show a way to reduce the problem of constructing seeded $s$-source non-malleable extractors to the problem of constructing non-malleable $(s+1)$-source extractors. Using the non-malleable $10$-source extractor with optimal error in \cite{CZ14}, we obtain a seeded non-malleable $9$-source extractor with optimal seed length, which in turn gives a $10$-source extractor for min-entropy $O(\log n)$. Previously the best known extractor for such min-entropy requires $O(\log \log n)$ sources \cite{Coh16b}.

Independent of our work, Cohen \cite{Cohen16} obtained similar results to (1) and the two-source extractor, except the dependence on $\e$ is $\log(1/\e)\polylog \log (1/\e)$ and the two-source extractor requires min-entropy $\log n \polylog \log n$.
\end{abstract}
\end{titlepage}

\section{Introduction}
Randomness extractors are fundamental objects in the study of pseudorandomness, a branch of modern theoretical computer science. Their motivations come from the need of uniform random bits in many applications, such as randomized algorithms, distributed computing, and cryptography, and the fact that natural random sources are almost always biased. Informally, randomness extractors transform imperfect random sources (whether naturally so or as a result of adversarial information leakage) into nearly uniform random bits, which can then be used in standard applications. Over the past decades randomness extractors have been extensively studied.

To model imperfect randomness, we use the by now standard model of a general weak random source with a certain amount of entropy. 

\begin{definition}
The \emph{min-entropy} of a random variable~$X$ is
\[ H_\infty(X)=\min_{x \in \supp(X)}\log_2(1/\Pr[X=x]).\]
For $X \in \zo^n$, we call $X$ an $(n,H_\infty(X))$-source, and we say $X$ has
\emph{entropy rate} $H_\infty(X)/n$.
\end{definition}

It is well known that by just having one weak source as input, no deterministic extractor can work for all $(n, k)$ sources even if $k=n-1$. Several ways are thus explored to get around this. One approach, introduced by Nisan and Zuckerman \cite{NisanZ96}, is to give the extractor an additional independent short uniform random seed. This results in the so called \emph{seeded extractors}.  

\begin{definition}(Seeded Extractor)\label{def:strongext}
A function $\Ext : \bits^n \times \bits^d \rightarrow \bits^m$ is  a \emph{$(k,\eps)$-extractor} if for every source $X$ with min-entropy $k$
and independent $Y$ which is uniform on $\zo^d$,
\[|\Ext(X, Y)-U_m | \leq \e.\]
If in addition we have $|(\Ext(X, Y), Y) - (U_m, Y)| \leq \e$ then we say it is a \emph{strong $(k,\eps)$-extractor}.
\end{definition}

One can show that seeded extractors with very good parameters exist for all $(n, k)$ sources, and with a long line of research their constructions are now close to optimal (e.g., \cite{LuRVW03, GuruswamiUV09, DvirW08, DvirKSS09}). Besides their original motivation, seeded extractors have found many other applications in theoretical computer science.

This paper, on the other hand, focuses on several other kinds of randomness extractors which have gained a lot of attention recently. The first one is extractors for \emph{independent sources}. Here, the extractor does not have any additional uniform random seed, but instead it is given as input more than one independent general weak random sources. The probabilistic method shows that deterministic extractors exist for just two independent $(n, k)$ sources with $k \geq \log n+O(1)$. In fact, with high probability a random function is such a two-source extractor. However, giving explicit constructions of such extractors turns out to be quite challenging.

The second kind of extractors we study here, focuses on the case where either the seed or the source is tampered with by an adversary. In this case, one useful and natural property to impose on the extractors is to ensure that the non-tampered output of the extractor is (close to) uniform even given the tampered output. This leads to a large class of generalized randomness extractors called \emph{non-malleable extractors}.

\begin{definition}[Tampering Funtion]
For any function $f:S \rightarrow S$, $f$ has a fixed point at $s \in S$ if $f(s)=s$. We say $f$ has no fixed points in  $T \subseteq S$, if $f(t) \neq t$ for all $t \in T$. We say $f$ has no fixed points if $f(s) \neq s$ for all $s \in S$.
\end{definition}

When the tampering acts on the seed of a seeded extractor, one obtains a generalization of strong seeded extractors called \emph{seeded non-malleable extractors}, originally introduced by Dodis and Wichs in \cite{DW09}.

\begin{definition}[Non-malleable extractor] A function $\snmExt:\{0,1\}^n \times \{ 0,1\}^d \rightarrow \{ 0,1\}^m$ is a seeded non-malleable extractor for min-entropy $k$ and error $\epsilon$ if the following holds : If $X$ is a  source on  $\{0,1\}^n$ with min-entropy $k$ and $\A : \{0,1\}^d \rightarrow \{0,1\}^d $ is an arbitrary tampering function with no fixed points, then
$$   |\snmExt(X,U_d) \scirc \snmExt(X,\A(U_d))  \scirc U_d- U_m \scirc  \snmExt(X,\A(U_d)) \scirc U_d | <\epsilon $$where $U_m$ is independent of $U_d$ and $X$.
\end{definition}

When the tampering acts on the sources in an independent source extractor, one obtains a generalization of independent source extractors called {seedless non-malleable extractors}, originally introduced by Cheraghchi and Guruswami \cite{CG14b}.  

\begin{definition}[Seedless Non-Malleable $C$-Source Extractor]\label{def:t2}
A function $\nmExt : (\{ 0,1\}^{n})^C \rightarrow \{ 0,1\}^m$ is a $(k, \e)$-seedless non-malleable extractor for $C$ independent sources, if it satisfies the following property: Let $X_1, \cdots, X_C$ be $C$ independent $(n, k)$ sources, and $f_1, \cdots, f_C : \zo^n \to \zo^n$ be $C$ arbitrary tampering functions such that there exists an $f_i$ with no fixed points, then  $$ |\nmExt(X_1, \cdots, X_C) \circ \nmExt(f_1(X_1), \cdots, f_C(X_2)) - U_m \circ \nmExt(f_1(X_1), \cdots, f_C(X_2))| < \epsilon.$$ Further, we say that the non-malleable extractor is strong if for \emph{every} $i$, we have that 

$$ |\nmExt(X_1, \cdots, X_C) \circ \nmExt(f_1(X_1), \cdots, f_C(X_2)) \circ X_i - U_m \circ \nmExt(f_1(X_1), \cdots, f_C(X_2)) \circ X_i| < \epsilon.$$
\end{definition}

We can also generalize the definition to handle more than one tampering functions.

\begin{definition}[Seeded $t$-Non-malleable extractor] A function $\snmExt:\{0,1\}^n \times \{ 0,1\}^d \rightarrow \{ 0,1\}^m$ is a seeded $t$-non-malleable extractor for min-entropy $k$ and error $\epsilon$ if the following holds : If $X$ is a  source on  $\{0,1\}^n$ with min-entropy $k$ and $\A_1, \cdots, \A_t : \{0,1\}^d \rightarrow \{0,1\}^d $ are $t$ arbitrary tampering functions with no fixed points, then
$$   |\snmExt(X,U_d) \scirc \{\snmExt(X,\A_i(U_d)), i \in [t]\} \scirc U_d- U_m \scirc  \{\snmExt(X,\A_i(U_d)), i \in [t]\} \scirc U_d| <\epsilon $$where $U_m$ is independent of $U_d$ and $X$.
\end{definition}

This definition can also be generalized to the case of seeded $t$-non-malleable extractor for more than one weak sources in the obvious way, and we omit the definition here.

As stated above, seeded non-malleable extractors were first introduced by Dodis and Wichs in \cite{DW09}, to study a cryptographic problem known as \emph{privacy amplification}. Although they seem to be irrelevant to independent source extractors, it turns out that these two kinds of extractors are closely related. Indeed, since the author's previous work \cite{Li12b, Li13a} which first established connections between seeded non-malleable extractors and independent source extractors, their connections have been demonstrated in several subsequent work. In particular, with other techniques, these connections have led to the recent breakthrough construction of two source extractors by Chattopadhyay and Zuckerman \cite{CZ15}. We now briefly review previous work below.
 
\paragraph{Independent source extractors.} The introduction of independent source extractors, as well as the first explicit construction of a two-source extractor appeared in \cite{ChorG88}, where Chor and Goldreich showed that the well known Lindsey's lemma gives an extractor for two independent $(n, k)$ sources with $k > n/2$. Since then there has been essentially no progress until Barak et. al\cite{BarakIW04} introduced new techniques in additive combinatorics into this problem, and constructed extractors for $O(1/\delta)$ independent $(n, \delta n)$ sources. Subsequently, a long line of fruitful results \cite{BarakIW04, BarakKSSW05, Raz05, Bourgain05, Rao06, BarakRSW06, Li11b, Li13a, Li13b, Li15, Cohen15} has introduced many new techniques and culminated in the three source extractor of exponentially small error for poly-logarithmic min-entropy by the author \cite{Li15}. In the case of two-source extractors, Bourgain \cite{Bourgain05} gave a construction that breaks the entropy rate $1/2$ barrier, and works for two independent $(n, 0.49n)$ sources. In a different work, Raz \cite{Raz05} gave an incomparable result of two source extractors which requires one source to have min-entropy larger than $n/2$, while the other source can have min-entropy $O(\log n)$. In a recent result, Chattopadhyay and Zuckerman \cite{CZ15} greatly improved the situation and gave the first explicit two-source extractor for $(n, k)$ sources with $k \geq \log^C n$ for some large enough constant $C$. Their construction only outputs one bit but this was later improved by the author to output almost all entropy \cite{Li16} and by Meka \cite{Mek:resil} to work for smaller min-entropy.

Very recently, there has been a new line of work focusing on constructing explicit independent source extractors for very small min-entropy (i.e., near logarithmic). Cohen and Schulman \cite{Coh16b} constructed extractors for $O(1/\delta)$ sources with min-entropy $\log^{1+\delta} n$. Chattopadhyay and Li \cite{CL16} improved this result to give an explicit extractor for $O(1)$ sources with min-entropy $\log n 2^{O(\sqrt{ \log \log n})}$, and this was subsequently improved by Cohen \cite{Coh16} to achieve a $5$-source extractor with the same entropy requirement. Finally, Ben-Aroya et. al \cite{BDT16} further improves this and achieves a two-source extractor for min-entropy $\log n 2^{O(\sqrt{ \log \log n})}$.


\paragraph{Seeded non-malleable extractors and privacy amplification.} As mentioned above, seeded non-malleable extractors were first introduced by Dodis and Wichs \cite{DW09} to study the question of privacy amplification with an active adversary, and they were later found to have close connections to independent source extractors. Thus, any progress in non-malleable extractors is likely to lead to progress in both the privacy amplification problem and the independent source extractor problem. 

Privacy amplification \cite{BennettBR88} is a basic problem in information theoretic cryptography, where two parties with local (non-shared) uniform random bits communicate through a public channel to convert a shared secret weak random source $\X$ into shared secret nearly uniform random bits. The communication channel is watched by an adversary Eve, who has unlimited computational power and tries to corrupt the protocol. Standard strong seeded extractors are enough to give very efficient protocols for this problem in the case where Eve is passive (i.e., can only see the messages but cannot change them). In the more complicated case where Eve is active (i.e., can arbitrarily change, delete and reorder messages), the goal is to design a protocol that uses as few number of interactions as possible, and outputs a shared uniform random string $\rr$ as long as possible (the difference between the length of the output and $H_{\infty}(\X)$ is called \emph{entropy loss}). The protocol is associated with a security parameter $s$, and ensures that if Eve is active, then the probability that Eve can successfully make the two parties output two different strings without being detected is at most $2^{-s}$. On the other hand, if Eve remains passive, then the two parties should achieve shared secret random bits that are $2^{-s}$-close to uniform. We refer the readers to \cite{DLWZ11} for a formal definition.

Much research has been devoted to this problem \cite{MW97,dkrs,DW09,RW03,KR09,ckor,DLWZ11,CRS11,Li12a,Li12b,Li15b}. It is known that when the entropy rate of $\X$ is large, i.e., bigger than $1/2$, there exist protocols that take only one round (e.g., \cite{MW97,dkrs}), albeit with quite large entropy loss. When the entropy rate of $\X$ is smaller than $1/2$, \cite{DW09} showed that any protocol has to take at least two rounds with entropy loss at least $O(s)$. Thus, the natural goal is to design a two-round protocol with such optimal entropy loss, for any possible security parameter (ideally up to $\Omega(k)$). However, all protocols before the work of \cite{DLWZ11} require $O(s)$ rounds or entropy loss $O(s^2)$. 

In \cite{DW09}, Dodis and Wichs further showed that two-round privacy amplification protocols with optimal entropy loss can be constructed using explicit seeded non-malleable extractors. Using the probabilistic method, they showed the existence of non-malleable extractors when $k>2m+2\log(1/\eps) + \log d + 6$ and $d>\log(n-k+1) + 2\log (1/\eps) + 5$. However, they were not able to give any explicit construction.\ The first explicit seeded non-malleable extractor was constructed in \cite{DLWZ11}, with subsequent improvements in \cite{CRS11,Li12a,DY12,Li12b}. Unfortunately all these constructions require min-entropy at least $0.49n$, and thus only give two-round privacy amplification protocols with optimal entropy loss for such min-entropy. Although, combined with other ideas, \cite{DLWZ11} also gives $\poly(1/\delta)$ round protocols with optimal entropy loss for min-entropy $k \geq \delta n$, any constant $\delta>0$. Subsequently, without improving on the non-malleable extractors, the author \cite{Li12b} gave a two-round protocol with optimal entropy loss for min-entropy $k \geq \delta n$, any constant $\delta>0$. Using a relaxation of non-malleable extractors called non-malleable condensers, the author \cite{Li15b} also obtained a two-round protocol with optimal entropy loss for min-entropy $k \geq C\log^2 n$, some constant $C>1$, as long as the security parameter $s$ satisfies $k \geq Cs^2$. 

The next improvement in non-malleable extractors appeared in \cite{CGL15}, where Chattopadhyay, Goyal and Li constructed explicit non-malleable extractors with error $\eps$, for min-entropy $k =\Omega( \log^2{(n/\epsilon)})$ and seed-length $d=O(\log^2(n/\epsilon))$. This gives an alternative protocol matching that of \cite{Li15b}. Further improvements were obtained by Cohen \cite{Coh15nm, Coh16a}, where he constructed non-malleable extractors with seed length $d= O(\log(n/\epsilon)\log((\log n)/\epsilon))$ and min-entropy $k = \Omega(\log(n/\epsilon)\log((\log n)/\epsilon))$;  seed-length $O(\log n)$ and min-entropy $k=n/(\log n)^{O(1)}$;  and seed length $d=O(\log n+ \log^3(1/\epsilon))$ and min-entropy $k=\Omega(d)$. However, none of these improves the privacy amplification protocols in \cite{Li15b}. 


Very recently, Chattopadhyay and Li \cite{CL16} obtained an improved non-malleable extractor with error $\eps$, for min-entropy $k =\log{(n/\epsilon)}2^{O(\sqrt{\log \log (n/\epsilon)})}$ and seed-length $d=\log{(n/\epsilon)}2^{O(\sqrt{\log \log (n/\epsilon)})}$, and min-entropy $k=O(\log n)$ and seed length $d=O(\log n)$ for error $\e \geq 2^{-\log^{1-\beta} n}$ for any constant $0< \beta<1$. Independently, Cohen \cite{Coh16} also obtained a non-malleable extractor with error $\eps$, for min-entropy $k =O(\log n) +\log(1/\epsilon)2^{O(\sqrt{\log \log (1/\epsilon)})}$ and seed-length $d =O(\log n) +\log(1/\epsilon)2^{O(\sqrt{\log \log (1/\epsilon)})}$. Both these constructions give two round privacy amplification protocols with optimal entropy loss, for security parameter $s$ up to $k/2^{O(\sqrt{\log k})}$.

\paragraph{Seedless non-malleable extractors and non-malleable codes.}
Seedless non-malleable extractors were first introduced by Cheraghchi and Guruswami \cite{CG14b}, in the context of non-malleable codes. Non-malleable codes, introduced by Dziembowski, Pietrzak and Wichs \cite{DPW10}, are a useful generalization of standard error correcting codes in the sense that they can handle a much larger class of attacks. Most notably, they can provide security guarantees even if the attacker can completely overwrite the codeword. Informally, a non-malleable code for a specific tampering family of tampering functions $\cal F$, consists of a randomized encoding function $E$ and a deterministic decoding function $D$, such that if a codeword $E(x)$ is modified into $f(E(x))$ by some function $f \in \cal F$, then the decoded message $x'=D(f(E(x)))$ is either the original message $x$, or a completely unrelated message. The formal definition is given in Section~\ref{sec:nmtext}. As shown in \cite{DPW10}, such non-malleable codes can be used in several applications in tamper-resilient cryptography. 

While it can be seen that even non-malleable codes cannot exist if $\cal F$ is completely unrestricted, it is also known to exist for many broad tampering families. One of the most natural tampering families, and the most well studied, is the so called \emph{split-state} model. Here, a $k$-bit message $x$ is encoded into $t$ parts of messages $y_1, \cdots, y_t$, each of length $n$. Now the adversary can arbitrarily tamper with each $y_i$ independently. In this case, the rate of the code is defined as $k/(tn)$. 

This model arises in many applications naturally, for example when the different parts of messages $y_1, \cdots, y_t$ are stored in different parts of memory. It can also be viewed as a kind of ``non-malleable secret sharing scheme". Clearly, the case of $t=1$ corresponds to unrestricted tampering functions, and cannot be handled by non-malleable codes. Thus the case of $t=2$ is the most useful and interesting setting. There has been a lot of work studying non-malleable codes in the $t$-split-state model. Since in this paper we focus on the information theoretic setting, we will only briefly review those previous work in the same setting.

The existence of non-malleable codes was first proved in \cite{DPW10}, and then Cheraghchi and Guruswami \cite{CG14a} improved this result to show that the optimal rate of non-malleable codes in the $2$-split-state model is $2$. The first explicit construction appears in \cite{DKO13}, where the authors constructed explicit non-malleable codes for $1$-bit messages in the split-state model. Subsequently, Aggarwal et. al \cite{ADL14} constructed the first explicit non-malleable code for $k$-bit messages. Their encoding has message length $n=O(k^7 \log^7 k)$. This was later improved by Aggarwal \cite{Agw14} to obtain $n=O(k^7)$. 

Cheraghchi and Guruswami \cite{CG14b} found a connection between non-malleable $t$-source extractors and non-malleable codes in the $t$-split state model. Their construction allows one to construct non-malleable codes in the $t$-split state model given sufficiently good non-malleable $t$-source extractors. However, they were not able to construct explicit non-malleable two-source extractors even for min-entropy $k=n$. Using this connection and techniques form additive combinatorics, Chattopadhyay and Zuckerman \cite{CZ14} constructed a non-malleable $10$-source extractor and a constant rate non-malleable code in the $10$-split-state model. In a subsequent work, Chattopadhyay, Goyal and Li \cite{CGL15} constructed the first explicit non-malleable two-source extractor for min-entropy $k=(1-\gamma)n$ with output $\Omega(k)$ and error $2^{-k^{\Omega(1)}}$, and used it to give an explicit non-malleable code in the $2$-split state model with rate $n^{-\Omega(1)}$. 

Finally, the work of  Aggarwal et.\ al \cite{ADKO15}, has a construction which ``achieves" a constant rate non-malleable code in the $2$-split-state model. However, recently the author found an error in their proof (we briefly discuss the error in Appendix~\ref{app:error}), and thus this result does not hold. Currently, only non-malleable codes of rate $n^{-\Omega(1)}$ can be deduced from their work.
 
\subsection{Our Results}
We obtain improved results in all of the above problems. First, we have the following theorem which gives improved constructions of seeded non-malleable extractors.

\BT \label{thm1}
There exists a constant $C>1$ such that for any $n, k \in \N$ and $0<\e<1$ with $k \geq C(\log n+\log \log(1/\e) \log(1/\e))$, there is an explicit strong seeded $(k, \e)$ non-malleable extractor $\zo^n \times \zo^d \to \zo^m$\ with $d=C(\log n+\log \log(1/\e) \log(1/\e))$ and $m \geq k/4$.
\ET

Combined with the protocol in \cite{DW09}, this gives the following theorem.

\BT
There exists a constant $0<\alpha<1$ such that for any $n, k \in \N$ and security parameter $s \leq \alpha k/\log k$, there is an explicit two-round privacy amplification protocol with entropy loss $O(\log n+s)$, in the presence of an active adversary.
\ET

Combined with the techniques in \cite{BDT16}, we obtain the following theorem which gives improved constructions of two-source extractors.

\BT 
For every constant $\e>0$ there exists a constant $c>1$ and an explicit two-source extractor $\Ext: (\zo^n)^2 \to \zo$ for min-entropy $k \geq c \log n \log \log n$, with error $\e$.
\ET

As a corollary, we obtain the following improved constructions of Ramsey graphs.

\BCR
For every large enough integer $N$ there exists a (strongly) explicit construction of a $K$-Ramsey graph on $N$ vertices with $K=(\log N)^{O(\log \log \log N)}$ 
\ECR

Next we give an improved construction of a non-malleable two-source extractor.

\BT 
There exists a constant $0< \gamma< 1$ and a non-malleable two-source extractor for $(n, (1-\gamma)n)$ sources with error $2^{-\Omega(n/\log n)}$ and output length $\Omega(n)$.
\ET

We give an algorithm to efficiently sample from the pre-image of this extractor, and together with the connection in \cite{CG14b}, we obtain the following theorem.

\BT 
For any $n \in \N$ there exists an explicit non-malleable code with efficient encoder/decoder in the $2$-split-state model with block length $2n$, rate  $\Omega(1/\log n)$ and error $=2^{-\Omega(n/\log n)}$.
\ET

Finally, we use the non-malleable $10$-source extractor in \cite{CZ14} to obtain the following theorem.

\BT 
For every constant $\e>0$ there exists a constant $c>1$ and an explicit ten-source extractor $\Ext: (\zo^n)^{10} \to \zo$ for min-entropy $k \geq c \log n$, with error $\e$.
\ET

\paragraph{Independent Work.} Independent of our work, and using different techniques, Cohen \cite{Cohen16} obtained similar results for seeded non-malleable extractors and two-source extractors. Specifically, he constructed seeded non-malleable extractors for seed length and min-entropy $O(\log n)+\log(1/\e)\polylog \log (1/\e)$, that outputs $\Omega(k)$ bits. He also constructed two-source extractors for min-entropy $\log n \polylog \log n$.

\subsection{Overview of The Constructions and Techniques}
Here we give a brief overview of our constructions and the techniques. Both our constructions of seeded non-malleable extractor and seedless non-malleable extractor follow the high level framework of recent constructions \cite{CGL15, Coh15nm, Coh16a, CL16, Coh16}. Specifically, we first obtain a small advice such that with high probability the untampered advice is different from the tampered version. The short size of the advice guarantees that even conditioned on the fixing of the advice, the seed and the source (or different sources) are still independent and have high min-entropy. We then use an improved correlation breaker with advice to obtain the output. Informally, given the advice, the correlation breaker does a series of computations using the inputs; and the output is guaranteed to be close to uniform given the tampered output, if the advice is different from the tampered advice. 

Take the seeded non-malleable extractor for example. It is well known that to achieve error $\e$, one can use an advice of length $O(\log(n/\e))$ (or even smaller, as shown in \cite{Coh16}), and length $\Omega(\log(1/\e))$ is necessary. Moreover, this only costs $O(\log(n/\e))$ bits in the seed and $O(\log(n/\e))$ entropy in the source. We now turn to the part of the correlation breaker with advice. This part is going to follow the recent developments in \cite{CL16, Coh16}, where (non-malleable) \emph{independence preserving mergers} are used to construct the correlation breaker with advice. Specifically, let us briefly recall what is done in \cite{CL16}. There, given the advice of length $L$, we first use an additional $O(\log(n/\e))$ bits to create a matrix of $L$ rows, such that each row corresponds to a bit in the advice and each is uniform (but may be correlated with other rows). The property guaranteed is that on the bit that is different in the advice and the tampered advice, the corresponding row in the matrix is uniform even conditioned on the corresponding row in the tampered version of the matrix. Then, using the rest of the bits from the seed, we merge the matrix into one final row, while keeping this independence.   

In \cite{CL16}, the construction first uses a basic merger, which uses $O(l \log(m/\e))$ random bits to merge a matrix of $l$ rows, each row having length $m$. Then, one chooses a particular $l$ and applies the basic merger to the initial matrix of $L$ rows, merging $l$ rows each time. This takes $\log L/\log l$ steps. Each step one needs to use fresh random bits. However, since there is also a tampered seed, if each time we use the same number of fresh random bits, then they may already contain no entropy given the previously leaked tampered seeds. Therefore, in \cite{CL16}, each time the number of fresh random bits used is at least twice as large as the number of random bits used in the previous step. This means the number of random bis needed is going to grow exponentially, and eventually we need $2^{O(\log L/\log l)}l \log(m/\e)$ random bits. A simple calculation shows that to minimize this quantity, we should choose $l$ such that $\log l =\sqrt{\log L}$ and this gives us $2^{O(\sqrt{\log L})} \log(m/\e)=2^{O(\sqrt{\log \log(n/\e)})} \log(m/\e)$ bits needed.

In this paper, we improve the merger in \cite{CL16}. From the above discussion, one can see that if somehow we can get around the bottleneck of doubling the length of the random bits used each time, then ideally we would just need $O(l \log L/\log l \log(m/\e))$ random bits. This quantity is minimized when $l$ is a constant (e.g., 2) and this gives us $O(\log L \log(m/\e))=O(\log \log (n/\e))\log(m/\e)$ random bits, which is much better than the previous one. How do we achieve this? Recall that previously the reason why we need to double the length of the random bits used each time, is that previously used bits from the tampered version can leak information about the current random bits of the untampered seed. If we can prevent this from happening, then we will be done. In other words, what we now need is to guarantee that each time the new random bits used in the seed is (close to) independent of  the random bits previously used in the tampered version. Our crucial observation is that this is exactly a ``look-ahead" property, and can be achieved by using alternating extraction.

This motivates the following construction. Let the source be $X$ and the seed be $Y$. After obtaining the advice, take a small slice $Y_1$ of $Y$ and use $Y_1$ to extract a small uniform output $Z$ from $X$. Use $Z$ and $Y$ (which still has a lot of entropy) to do an alternating extraction and output $\log L+1$ random variables $R_i$. One can show that conditioned on the fixing of $Z$, these random variables are all deterministic functions of $Y$, and each $R_i$ is close to uniform conditioned on the previous ones and the previous tampered ones (i.e., they satisfy the look-ahead property). Now, we can use $R_1$ and $X$ (which, again, still has a lot of entropy) to create the initial matrix of $L$ rows, and then subsequently each time use a new $R_i$ to merge this matrix. 

The above construction almost achieves what we want, except one problem. The problem is that the basic merger, which uses alternating extraction itself, only outputs say $0.2m$ bits if originally each row has $m$ bits (think of the non-malleable extractor case, which can output at most $k/2$ bits if the min-entropy is $k$). Thus, if we simply repeat the merging step for $\log L$ steps, then the length of the output will decrease to $2^{-O(\log L)}m$; and for this to be meaningful we would need $m \geq 2^{O(\log L)}$, which would make $m$ and also the min-entropy $k$ become at least $\poly(L)=\polylog(n/\e)$. This is too large for our goal. Thus, we modify this construction so that we can compensate for the loss of output length each time. Specifically, after obtaining the advice, we first take a small slice $Y_1$ of $Y$ and use $Y_1$ to extract a small uniform output $Z$ from $X$. Note that conditioned on the fixing of $Y_1$, $Z$ is a deterministic function of $X$. Now we take a slightly larger slice $Y_2$ of $Y$, and a slice $Z_2$ of $Z$. Note that given $(Y_1, Y_2)$, $Y$ still has a lot of entropy. Similarly, given $(Y_1, Z_2)$, $Z$ still has a lot of entropy. We will now first use $Z_2$ and $Y$ to do an alternating extraction and output $2\log L+1$ random variables $R_i$. We will also use $Y_2$ and $Z$ to do an alternating extraction and output $\log L+1$ random variables $S_i$. One can show that conditioned on $(Y_1, Z)$, all the $R_i$ are deterministic functions of $Y$, and satisfy the look-ahead property. Similarly, conditioned on $(Y_1, Y_2)$,  all the $S_i$ are deterministic functions of $X$, and satisfy the look-ahead property. We now use $S_0$ and $R_0$ (the first blocks in the sequences) to obtain the initial matrix, which conditioned on the fixing of $R_0$ is a deterministic function of $S_0$. Then, we repeat the merging for $\log L$ steps. Each step we will use two $R_i$'s and one $S_i$. Consider a particular step $i$. We first use $R_{2i-1}$ to merge the matrix, reducing the number of rows to a half. Note that conditioned on the fixing of $R_{2i-1}$, the output is a deterministic function of $S_{i-1}$. We then use each row of the output as a seed to extract from $R_{2i}$. Now conditioned on the previous matrix, the new output is a deterministic function of $R_{2i}$. Finally, we use each row of the new output as a seed to extract from $S_{i}$. Conditioned on the fixing of $R_{2i}$, the output becomes a deterministic function of $S_{i}$, and by choosing the length of each $S_{i}$ to be larger than $2m$ we can restore the length of each row in the matrix to $m$. This whole process still preserves the independence between the matrix and the tampered version of the matrix. We can thus repeat the process until we obtain the final output. Note that for all the alternating extraction, we can control the length of $Z$ and $S_i$, so that the number of random bits used is smaller than $O(\log(n/\e))$. We also need to set $\e$ to be slightly smaller than the error we want to achieve. Careful calculations show that we can achieve the seed length and entropy requirement in Theorem~\ref{thm1}. By setting the parameters correctly, we can also ensure that the whole process described above does not consume much entropy, thus we can use the final output to extract from the original source and output $\Omega(k)$ bits. 

The non-malleable two-source extractor follows essentially the same construction, except we now know that both sources already have min-entropy $(1-\gamma)n$. Thus, we can afford to set the error parameter to be $2^{-\Omega(n/\log n)}$.

\paragraph{Efficient sampling.}The above non-malleable two-source extractor implies a non-malleable code in the $2$-split-state model with rate $\Omega(1/\log n)$. However, to obtain an efficient encoder, we need to find a way to efficiently sample uniformly from the pre-image of any given output.  Since the construction of the non-malleable two-source extractor is complicated and involves  multi steps of alternating extraction etc., it appears that the sampling procedure may also be complicated. Indeed, in \cite{CGL15} the sampling procedure consists of a series of carefully designed steps to ``invert" each intermediate extraction step. Here, we show that in fact we can significantly simplify the sampling procedure. In fact, we are going to treat most of the details in the construction of the non-malleable two-source extractor as a black box, and all we need are two ingredients from \cite{CGL15}: First, a seeded extractor $\iext: \bits^n \times \bits^d \to \bits^m$ with $d=O(\log(n/\e))$ and $m=\Omega(d)$, such that for any fixed output $s$ and any fixed seed $r$, one can efficiently uniformly sample from the pre-image (this is because for any fixed seed, the output is a linear function of the input source), and the pre-image always has the same size. Second, to obtain the advice, first we take a small slice $X_1$ of the source $X$, and a small slice $Y_1$ of the source $Y$. Both slices have size $3\gamma n$ (assuming both sources have min-entropy $(1-\gamma)n$). We take the inner product of $X_1$ and $Y_1$, and use the output to sample $\Omega(n/\log n)$ coordinates from the Reed-Solomon encodings of both the rest part of $X$ and the rest part of $Y$. The advice $\alpha$ is obtained by concatenating $X_1$, $Y_1$ and the sampled coordinates. Now we slightly modify the non-malleable two-source extractor in the following way. We will take two other slices $Y_2$ and $Y_3$ of $Y$, with the guarantee that each has high min-entropy conditioned on previously leaked information, and the total length of $(Y_1, Y_2, Y_3)$ is less than $n/2$ (but still $\Omega(n)$). Similarly we take another slice $X_2$ of $X$, which has high min-entropy conditioned on previously leaked information, and the total length of $(X_1, X_2)$ is less than $n/2$ (but still $\Omega(n)$). Given the advice, we use $(X_2, Y_2)$ to run the non-malleable two source extractor described above, and obtain an output $V$. We then compute the final output $W=\iext(Y_3, V)$. The non-malleable two-source extractor guarantees that $V$ is close to uniform given the tampered version, and this will be preserved in $W$. 

Given any output $W$, we now briefly describe how to efficiently uniformly sample from the pre-image. We first uniformly generate $(X_1, Y_1, X_2, Y_2)$ and the advice $\alpha$. From these things we can compute the output $V$. Note that here we are treating the details in the construction of the non-malleable two-source extractor as a black box. Now, given $V$ and $W$, by the property of $\iext$ we can efficiently sample $Y_3$, and the pre-image always has the same size. Finally, we need to sample the rest parts of $X$ and $Y$, given the variables we have obtained and $\alpha$. For this step, we note that once we have $(X_1, Y_1)$, we know the coordinates of the Reed-Solomon codes that we sampled, and these give us a system of linear equations. Note that we have at least $n/2$ free variables in both $X$ and $Y$, thus by setting the length of the advice appropriately (which is $\Omega(n)$) we can ensure that there are more variables in the system of equations than the number of equations. Therefore we can efficiently sample the pre-image by inverting the system of linear equations. Further note that the encoding matrix of the Reed-Solomon code has the property that regardless of the positions of the coordinates, as long as the number of sampled coordinates is the same, the encoding matrix always has the same rank. Thus the pre-image also has the same size regardless of the positions of the coordinates sampled. Therefore, altogether we can efficiently uniformly sample from the pre-image.

\paragraph{Independent source extractor.}  A corollary of the work of  Ben-Aroya et. al \cite{BDT16} is that if one can construct seeded $t$-non-malleable extractor for some constant $t$ with error $\e$, seed length and min-entropy $O(\log(n/\e))$, then one also gets an explicit two-source extractor for min-entropy $O(\log n)$. The two-source extractor outputs one bit with any constant error. In this paper we show that we can reduce the task of constructing such seeded non-malleable extractor to the task of constructing non-malleable two-source extractors for $(n, (1-\gamma)n)$ sources with error $2^{-\Omega(n)}$, where $\gamma$ is any constant.

To see this, suppose we have such a non-malleable two-source extractor, then we can construct a seeded non-malleable extractor roughly as follows. Let the seed be $Y$ and the source be $X$. First, we can take a small slice of $Y$ and use it as a seed in an extractor, to convert $X$ into a close to uniform string. Let the result be $\bar{X}$. Then, as usual, we obtain an advice $\alpha$ such that $\alpha \neq \alpha'$ with high probability, where $\alpha'$ is the tampered version of $\alpha$. Now, we take a small slice $Y_2$ of $Y$, and a small slice $X_2$ of $\bar{X}$, with the guarantee that both slices have entropy rate $>1/2$. We take the inner product of $(X_2, Y_2)$, and use this output as an extractor to convert both $\bar{X}$ and $Y$ back into nearly uniform strings (the reason why we can do this is that the inner product is a two-source extractor strong in both sources). Let the outputs be $\tilde{X}$ and $\tilde{Y}$. We can now append $\alpha$ to both $\tilde{X}$ and $\tilde{Y}$. By setting the lengths appropriately we obtain two independent (conditioned on the fixing of previous random variables) $(m, (1-\gamma)m)$ sources, where $m=O(\log(n/\e))$ as long as both $X$ and $Y$ have min-entropy at least $C\log(n/\e)$ for some constant $C>1$. We know that with high probability both sources will be different than their tampered version, thus we can now apply the non-malleable two-source extractor to get an output with error $\e$.

The above construction is just for one tampering function, but we can use an argument similar to that used in \cite{Li13b, Cohen15} to gradually increase the resilience, until eventually the extractor works for $t$ tampering functions. This puts an $O(t^2)$ factor on the seed length and entropy requirement, which is still a constant if $t$ is a constant.

Clearly, the approach described above works not just for non-malleable extractors with optimal error, but works for any non-malleable extractor. Thus our non-malleable two-source extractor directly implies a two-source extractor for $(n, O(\log n \log \log n))$ sources. The approach also extends naturally to the case of non-malleable $(s+1)$-source extractor, which would give a seeded non-malleable extractor for $s$ independent sources. Thus, we can use the non-malleable $10$-source extractor with optimal error in \cite{CZ14}, which gives a seeded non-malleable extractor for $9$ independent sources. Together with the construction in \cite{BDT16} this gives an explicit extractor for $10$ independent $(n, O(\log n))$ sources, which outputs one bit with any constant error.\\

\noindent{\bf Organization.}
The rest of the paper is organized as follows.\ We give some preliminaries in Section~\ref{sec:prelim}. We then define alternating extraction in Section~\ref{sec:alt}, and non-malleable independence preserving merger in Section~\ref{sec:nmipm}. In Section~\ref{sec:advcb} we construct the new correlation breaker with advice. In Section~\ref{sec:snmext} we present the seeded non-malleable extractor. In Section~\ref{sec:nmtext} we present non-malleable two-source extractors and non-malleable codes in the two-split-state model. Section~\ref{sec:indext} gives constructions of $t$-non-malleable extractors and applications to independent source extractors. Finally we conclude with some discussions and open problems in \sectionref{sec:conc}.

\input{prelim.tex}

\input{reduction.tex}

\input{ext.tex}

\section{Conclusions and Open Problems}\label{sec:conc}
Previous work in the literature have established connections between seeded non-malleable extractors and two-source extractors, and connections between non-malleable two-source (or multi-source) extractors and non-malleable codes in the split-state model. In this paper we further established connections between seeded non-malleable extractors and non-malleable two-source extractors. Thus, all these four objects are closely related to each other. Using improved independence preserving mergers, we give improved constructions of seeded non-malleable extractors, two-source extractors, non-malleable two-source extractors and non-malleable codes in the two-split-state model. These constructions are quite close to optimal (in terms of the entropy requirement). Thus, the obvious open problem is to achieve optimal constructions for all of them, i.e., seeded non-malleable extractor with seed length and entropy $O(\log(n/\e))$, non-malleable two-source extractor for entropy $(1-\gamma)n$ with error $2^{-\Omega(n)}$ and output length $\Omega(n)$. In turn, these will give explicit two-source extractors for $O(\log n)$ min-entropy (with one bit output and any constant error), and constant-rate non-malleable codes in the two-split-state model.

On the other hand, all recent constructions of two-source extractors follow the framework of \cite{CZ15}, and thus the error is either $1/\poly(n)$ or any constant. So far, negligible error can only be achieved by using three sources \cite{Li15}, or two-sources when the min-entropy is at least $0.49n$ \cite{Bourgain05}. Constructing two-source extractors with smaller error, for smaller min-entropy is an interesting open problem, and seems to require new ideas.
\bibliographystyle{alpha}

\bibliography{refs}

\appendix 
\section{The error in \cite{ADKO15}} \label{app:error}
The error in \cite{ADKO15} appears in the proof of Theorem 26 (Section 5.3), which reduces two look-ahead tampering to a $t$-split tampering. Specifically, to prove Equation (9) there one needs to argue about the quantity $H_{\infty}(L_i|\mathsf{Var}_i)=H_{\infty}(L_i | Z_1, \cdots, Z_{i-1})$. The claim is that $H_{\infty}(L_i | Z_1, \cdots, Z_{i-1}) \geq n/2$ because $L_i$ is a uniform string on $n$ bits, and the size of $(Z_1, \cdots, Z_{i-1})$ is at most $n/2$. However, this is not true. The only thing one can make sure is that the size of $(h_1(U^{(1)}, Z_1), \cdots, h_{i-1}(U^{(i-1)}, Z_{i-1}))$ is at most $n/2$, as written in the proof. But these are functions of $(Z_1, \cdots, Z_{i-1})$ and only output partial information. By examining the definition of $\{Z_i\}$, one can see that each $Z_i$ has $m \cdot 2^m$ bits, thus the size of $(Z_1, \cdots, Z_{i-1})$ can be up to $tm2^m$. Therefore, in order to make sure this is less than $n/2$, one needs $n \geq 2tm2^m$ in the theorem, rather than $n \geq 2tm$ as currently written.

We note that at this time, it is still not clear whether the proof can be fixed.
\end{document}

%% file: prelim.tex
\section{Preliminaries} \label{sec:prelim}
We often use capital letters for random variables and corresponding small letters for their instantiations. Let $|S|$ denote the cardinality of the set~$S$. For $\ell$ a positive integer,
$U_\ell$ denotes the uniform distribution on $\zo^\ell$. When used as a component in a vector, each $U_\ell$ is assumed independent of the other components.
All logarithms are to the base 2.

\subsection{Probability distributions}
\begin{definition} [statistical distance]Let $W$ and $Z$ be two distributions on
a set $S$. Their \emph{statistical distance} (variation distance) is
\begin{align*}
\Delta(W,Z) \eqdef \max_{T \subseteq S}(|W(T) - Z(T)|) = \frac{1}{2}
\sum_{s \in S}|W(s)-Z(s)|.
\end{align*}
\end{definition}

We say $W$ is $\eps$-close to $Z$, denoted $W \approx_\eps Z$, if $\Delta(W,Z) \leq \eps$.
For a distribution $D$ on a set $S$ and a function $h:S \to T$, let $h(D)$ denote the distribution on $T$ induced by choosing $x$ according to $D$ and outputting $h(x)$.

\BL \label{lem:sdis}
For any function $\alpha$ and two random variables $A, B$, we have $\Delta(\alpha(A), \alpha(B)) \leq \Delta(A, B)$.
\EL

\subsection{Somewhere Random Sources and Extractors}

\begin{definition} [Somewhere Random sources] \label{def:SR} A source $X=(X_1, \cdots, X_t)$ is $(t \times r)$
  \emph{somewhere-random} (SR-source for short) if each $X_i$ takes values in $\bits^r$ and there is an $i$ such that $X_i$ is uniformly distributed.
\end{definition}

\begin{definition}(Seeded Extractor)\label{def:strongext}
A function $\Ext : \bits^n \times \bits^d \rightarrow \bits^m$ is  a \emph{strong $(k,\eps)$-extractor} if for every source $X$ with min-entropy $k$
and independent $Y$ which is uniform on $\zo^d$,
\[ (\Ext(X, Y), Y) \approx_\eps (U_m, Y).\]
\end{definition}

\subsection{Average conditional min-entropy}
\label{avgcase}


\begin{definition}
The \emph{average conditional min-entropy} is defined as
\[ \thinf(X|W)= - \log \left (\expect_{w \leftarrow W} \left [ \max_x \Pr[X=x|W=w] \right ] \right )
= - \log \left (\expect_{w \leftarrow W} \left [2^{-\hinf(X|W=w)} \right ] \right ).
\]
\end{definition}

\begin{lemma} [\cite{dors}]
\label{entropies}
For any $s > 0$,
$\Pr_{w \leftarrow W} [\hinf(X|W=w) \geq \thinf(X|W) - s] \geq 1-2^{-s}$.
\end{lemma}

\BL [\cite{dors}] \label{lem:amentropy}
If a random variable $B$ has at most $2^{\ell}$ possible values, then $\thinf(A|B) \geq \hinf(A)-\ell$.
\EL

\subsection{Prerequisites from previous work}

Sometimes it is convenient to talk about average case seeded extractors, where the source $X$ has average conditional min-entropy $\thinf(X|Z) \geq k$ and the output of the extractor should be uniform given $Z$ as well. The following lemma is proved in \cite{dors}.

\BL \cite{dors} \label{lem:avext}
For any $\delta>0$, if $\Ext$ is a $(k, \e)$ extractor then it is also a $(k+\log(1/\delta), \e+\delta)$ average case extractor.
\EL

For a strong seeded extractor with optimal parameters, we use the following extractor constructed in \cite{GuruswamiUV09}.

\BT [\cite{GuruswamiUV09}] \label{thm:optext} 
For every constant $\alpha>0$, and all positive integers $n,k$ and any $\e>0$, there is an explicit construction of a strong $(k,\e)$-extractor $\Ext: \bits^n \times \bits^d \to \bits^m$ with $d=O(\log n +\log (1/\e))$ and $m \geq (1-\alpha) k$. In addition, for any $\e> 2^{-k/3}$ this gives a strong $(k, \e)$ average case extractor with $m \geq k/2$.
\ET

\BT [\cite{ChorG88}] \label{thm:ip}
For every $0<m< n$ there is an explicit two-source extractor $\bip: \bits^n \times \bits^n \to \bits^m$ based on the inner product function, such that if $X, Y$ are two independent $(n, k_1)$ and $(n, k_2)$ sources respectively, then

\[(\bip(X, Y), X) \approx_{\e} (U_m, X) \text{ and } (\bip(X, Y), Y) \approx_{\e} (U_m, Y),\]
where $\e=2^{-\frac{k_1+k_2-n-m-1}{2}}.$
\ET

We need the following explicit construction of seedless non-malleable extractors in \cite{CZ14}.

\BT
There exists a constant $\delta>0$ and an explicit $(k, \e)$-seedless non-malleable extractor for $10$ independent sources $\czext: (\bits^n)^{10} \to \bits^m$ with $k=(1-\delta)n$, $\e=2^{-\Omega(n)}$ and $m=\Omega(k)$.
\ET

The following standard lemma about conditional min-entropy is implicit in \cite{NisanZ96} and explicit in \cite{MW97}.

\begin{lemma}[\cite{MW97}] \label{lem:condition} 
Let $X$ and $Y$ be random variables and let ${\cal Y}$ denote the range of $Y$. Then for all $\e>0$, one has
\[\Pr_Y \left [ H_{\infty}(X|Y=y) \geq H_{\infty}(X)-\log|{\cal Y}|-\log \left( \frac{1}{\e} \right )\right ] \geq 1-\e.\]
\end{lemma}

We also need the following lemma.

\BL \label{lem:jerror}\cite{Li13b}
Let $(X, Y)$ be a joint distribution such that $X$ has range $\calX$ and $Y$ has range $\calY$. Assume that there is another random variable $X'$ with the same range as $X$ such that $|X-X'| = \e$. Then there exists a joint distribution $(X', Y)$ such that $|(X, Y)-(X', Y)| = \e$.
\EL

\BL \label{lem:derror} \cite{BarakIW04}
Assume that $Y_1, Y_2, \cdots, Y_t$ are independent random variables over $\bits^n$ such that for any $i, 1 \leq i \leq t$, we have $|Y_i-U_n| \leq \e$. Let $Z=\oplus_{i=1}^{t} Y_i$. Then $|Z-U_n| \leq \e^t$.
\EL

%% file: reduction.tex
\section{Alternating Extraction}\label{sec:alt}
An important ingredient in our construction is the following alternating extraction protocol, which was first introduced in \cite{DP07}, and then used a lot in constructions related to extractors (e.g., \cite{DW09, Li13b}).

\begin{figure}[htb]
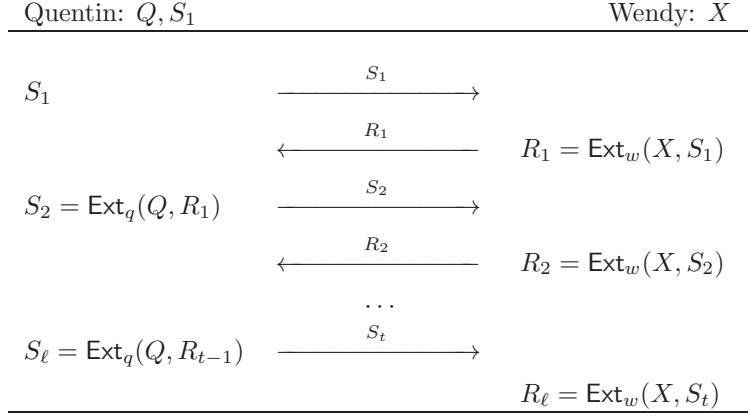

\begin{center}
\begin{small}
\begin{tabular}{l c l}
Quentin:  $Q, S_1$ & &~~~~~~~~~~Wendy: $X$ \\

\hline\\
$S_1$ & $\llrightarrow[\rule{2.5cm}{0cm}]{S_1}{} $ & \\
& $\llleftarrow[\rule{2.5cm}{0cm}]{R_1}{} $ & $R_1=\Ext_w(X, S_1)$ \\
$S_2=\Ext_q(Q, R_1)$ & $\llrightarrow[\rule{2.5cm}{0cm}]{S_2}{} $ & \\
& $\llleftarrow[\rule{2.5cm}{0cm}]{R_2}{} $ & $R_2=\Ext_w(X, S_2)$ \\
& $\cdots$ & \\
$S_\ell=\Ext_q(Q, R_{t-1})$ & $\llrightarrow[\rule{2.5cm}{0cm}]{S_t}{} $ & \\
& & $R_\ell=\Ext_w(X, S_t)$ \\
\hline
\end{tabular}
\end{small}
\caption{\label{fig:altext}
Alternating Extraction.
}
\end{center}
\end{figure}

\textbf{Alternating Extraction.} Assume that we have two parties, Quentin and Wendy. Quentin has a source $Q$,  Wendy has a source $W$. Also assume that Quentin has a uniform random seed $S_1$ (which may be correlated with $Q$). Suppose that $(Q, S_1)$ is kept secret from Wendy and $W$ is kept secret from Quentin.  Let $\Ext_q$, $\Ext_w$ be strong seeded extractors with optimal parameters, such as that in \theoremref{thm:optext}. Let $s$ be an integer parameter for the protocol. For some integer parameter $\ell>0$, the \emph{alternating extraction protocol} is an interactive process between Quentin and Wendy that runs in $\ell$ steps. 

In the first step, Quentin sends $S_1$ to Wendy, Wendy computes $R_1=\Ext_w(W, S_1)$. She sends $R_1$ to Quentin and Quentin computes $S_2=\Ext_q(Q, R_1)$. In this step $R_1, S_2$ each outputs $s$ bits. In each subsequent step $i$, Quentin sends $S_i$ to Wendy, Wendy computes $R_i=\Ext_w(W, S_i)$. She replies $R_i$ to Quentin and Quentin computes $S_{i+1}=\Ext_q(Q, R_i)$. In step $i$, $R_i, S_{i+1}$ each outputs $s$ bits. Therefore, this process produces the following sequence: 

\begin{align*}
S_1, R_1=\Ext_w(W, S_1), S_2=\Ext_q(Q, R_1), \cdots, S_\ell=\Ext_q(Q, R_{\ell-1}), R_\ell=\Ext_w(W, S_\ell).
\end{align*}

\textbf{Look-Ahead Extractor.} Now we can define our look-ahead extractor. Let $Y=(Q, S_1)$ be a seed, the look-ahead extractor is defined as 

\[\laext(W,  Y)=\laext(W, (Q, S_1)) \eqdef R_1, \cdots, R_{\ell}.\]

The following lemma is a special case of Lemma 6.5 in \cite{CGL15}. 

\begin{lemma}\label{altext} Let $W$ be an  $(n_{w},k_{w})$-source and $W'$ be a random variable on $\{ 0,1\}^{n_w}$ that is arbitrarily correlated with $W$. Let $Y=(Q, S_1)$ such that $Q$ is a $(n_q,k_q)$-source,  $S_1$ is a uniform string on $s$ bits, and $Y' = (Q', S'_1)$ be a random variable arbitrarily correlated with $Y$, where $Q'$ and $S'_1$ are random variables on $n_q$ bits and $s$ bits respectively. Let $\Ext_q,\Ext_w$ be strong seeded extractors that extract $s$ bits from sources with min-entropy $k$ with error $\epsilon$ and seed length $s$. Suppose $(Y, Y')$ is independent of $(W,W')$, and $k_w,k_q \ge k+ 2 \ell s+2 \log(\frac{1}{\epsilon})$. Let $\laext$ be the look-ahead extractor defined above using $\Ext_q,\Ext_w$, and $(R_1, \cdots, R_{\ell})=\laext(W, Y)$, $(R'_{1}, \cdots, R'_{\ell})=\laext(W', Y')$. Then for any $0 \leq j \leq \ell-1$, we have

\[(Y, Y',  \{R_{1}, R'_1, \cdots, R_{j}, R'_j\}, R_{j+1}) \approx_{\e_1} (Y, Y',  \{R_{1}, R'_1, \cdots, R_{j}, R'_j\}, U_s),\]
where $\e_1=O(\ell \e)$.
\end{lemma}

\section{Non-Malleable Independence Preserving Merger}\label{sec:nmipm}
We now describe the notion of \emph{non-malleable independence preserving merger}, introduced in \cite{CL16} based on the notion of independence preserving merger introduced in \cite{Coh16b}. For simplicity we assume here we only have one adversary, which will be enough for our applications.

\BD A $(L, d', \eps)$-$\nipm: \zo^{Lm} \times \zo^d \rightarrow \zo^{m_1}$ satisfies the following property.  Suppose
\begin{itemize}
\item $\X,\X'$ are random variables, each supported on boolean $L\times m$ matrices s.t for any $i \in [L]$, $\X_i = \U_m$,
\item $\{\Y,\Y'\}$ is independent of $\{ \X,\X'\}$, s.t $\Y,\Y'$ are each supported on $\zo^{d}$ and $H_{\infty}(\Y) \ge d'$,
\item there exists an $h \in [L]$ such that $(\X_h,\X'_h)=(\U_m,\X'_h)$,
\end{itemize}
then 
\begin{align*}
|(L,d', \eps)\text{-}\nipm((\X,\Y), (L,d', \eps)\text{-}\nipm(\X', \Y') -\U_{m_1},  (L,d', \eps)\text{-}\nipm(\X', \Y')| \le \epsilon.
\end{align*}
\ED

We have the following construction and theorem.

\textbf{$L$-Alternating Extraction} We extend the previous alternating extraction protocol by letting Quentin have access to $L$ sources $Q_1,\ldots,Q_L$ (instead of just $Q$) which have the same length.  Now in the $i$'th round of the protocol, he uses $Q_i$ to produce the r.v $S_i=\Ext_{q}(Q_i,R_i)$. More formally, the following sequence of r.v's is  generated: $ S_1, R_1 = \Ext_{w}(W,S_1), S_2 = \Ext_{q}(Q_2, R_1),\ldots,R_{L-1}=\Ext_{w}(W, S_{\ell-1}),S_{L} = \Ext_{q}(Q_{L},R_{L-1})$. 

The $\nipm$ is now constructed as follows. Let $S_1$ be a slice of $\X_1$ with length $O(\log(d/\eps))$, then run the $L$-alternating extraction described above with $(Q_1,\ldots,Q_L)=(\X_1, \ldots, \X_L)$ and $W=\Y$. Finally output $S_L$.

\begin{theorem}[\cite{CL16}] \label{thm:nipm} 
There exists a constant $c>0$ such that for all integers $m,d, d', L>0$  and any $\epsilon>0$,  with $m \ge 4c L \log(d/\epsilon)$, $d' \ge 4c L \log(m/\epsilon)$, the above construction $\nipm:(\zo^{m})^{\ell} \times \zo^d \rightarrow \zo^{m_1}$ has output length $m_1 \geq 0.2 m$, such that if the following conditions hold:

\begin{itemize}
\item $\X,\X'$ are random variables, each supported on boolean $L\times m$ matrices s.t for any $i \in [L]$, $\X_i = \U_m$,
\item $\{\Y,\Y'\}$ is independent of $\{ \X,\X'\}$, s.t $\Y,\Y'$ are each supported on $\zo^{d}$ and $H_{\infty}(\Y) \ge d'$,
\item there exists an $h \in [L]$ such that $(\X_h,\X'_h)=(\U_m,\X'_h)$,
\end{itemize}
then 
\begin{align*}
| \nipm((\X, \Y), \nipm((\X', \Y'), \Y, \Y' -\U_{m_1}, \nipm((\X', \Y'), \Y,\Y'| \le L \epsilon.
\end{align*}
\end{theorem}

\section{Correlation Breaker with Advice}\label{sec:advcb}
We now use the non-malleable independence preserving merger to construct an improved correlation breaker with advice. A correlation breaker, as its name suggests, uses independent randomness to break the correlations between several correlated random variables. A prototype correlation breaker was first constructed implicitly in the author's work \cite{Li13b}, and then later strengthened and formally defined in \cite{Cohen15}. A correlation breaker with advice additionally uses some string as an advice. This object was first introduced and used without its name in \cite{CGL15}, and then explicitly defined in \cite{Coh15nm}. We have the following definition.

\BD [Correlation breaker with advice] A function

\[\acb: \bits^n \times \bits^d \times \bits^a \to \bits^m\] is called a $(k, \eps)$-correlation breaker with advice if the following holds. Let $Y, Y'$ be $d$-bit
random variables such that $Y$ is uniform. Let $X, X'$ be $n$-bit random variables with $H_{\infty}(X) \geq k$, such that $(X, X')$ is independent of $(Y, Y')$. Then, for any pair of distinct $a$-bit strings $\alpha, \alpha'$,

\[(\acb(X,Y,\alpha), \acb(X',Y',\alpha')) \approx_{\eps} (U,\acb(X',Y',\alpha')).\] 
In addition, we say that $\acb$ is strong if

\[(\acb(X,Y,\alpha), \acb(X',Y',\alpha'), Y, Y') \approx_{\eps} (U,\acb(X',Y',\alpha'), Y, Y').\] 
\ED

For our construction we need the following flip-flop extraction scheme. The flip-flop function was constructed by Cohen \cite{Cohen15} using alternating extraction, based on a previous similar construction of the author \cite{Li13b}. Subsequently, it was used in the construction of non-malleable extractors by Chattopadhyay, Goyal and Li \cite{CGL15}. The flip-flop function is a basic version of correlation breaker, and (informally) uses an independent source $\X$ to break the correlation between two r.v's $\Y$ and $\Y'$, given an advice bit. We now describe this more formally.

\begin{theorem}[\cite{Cohen15,CGL15}]\label{flip} There exists a constant $c_{\ref{flip}}$ such that for all $n>0$ and any $\epsilon>0$, there exists an explicit function $\flip:\zo^n \times \zo^d \rightarrow \zo^m$, $m=0.4 k$,   satisfying the following: Let $\X$ be an $(n,k)$-source, and $\X'$ be a random variable on $n$ bits arbitrarily correlated with $\X$. Let $\Y$ be an independent uniform seed on $d$ bits, and $\Y'$ be a random variable on $d$ bits arbitrarily correlated with $\Y$. Suppose $(\X, \X'$) is independent of $(\Y, \Y')$.  If $k,d \ge C_{\ref{flip}}\log(n/\epsilon)$,  then for any bit $b$, 
$$|\flip(\X,\Y,b), \Y,\Y' - \U_m,\Y,\Y'| \le \epsilon.$$
Furthermore, for any bits $b, b'$ with $b \neq b'$,  
$$|\flip(\X,\Y,b),\flip(\X',\Y',b'),\Y,\Y' - \U_m,\flip(\X',\Y',b'),\Y,\Y'| \le \epsilon.$$
\end{theorem}

We construct a correlation breaker such that $X, X', Y, Y'$ are all on $d$ bits such that $H_{\infty}(X) \geq 0.9d$ and $H_{\infty}(Y) \geq 0.9d$. Using the above ingredients, our construction of the correlation breaker with advice is given below. For simplicity, when we say a strong seeded extractor for min-entropy $k$, we mean a strong average case seeded extractor for average conditional min-entropy $k$.  

\begin{itemize}
\item Fix an error parameter $\e'$ to be chosen later. Let $s$ be an integer such that $s \geq \max\{c \log(d/\e'), 8c \log (3s/\e')\}$ where $c$ is the maximum of the hidden constant in the seed length of the optimal seeded extractor in Theorem~\ref{thm:optext}, and the two constants $c, c_{\ref{flip}}$ in Theorem~\ref{thm:nipm} and Theorem~\ref{flip}. 

\item Let $\Ext$ be a strong seeded extractor which uses $r=c \log (3s/\e')$ random bits to extract from an $(3s, 2c \log (3s/\e'))$ source and outputs $r=c \log (3s/\e')$ bits with error $\e'$, from Theorem~\ref{thm:optext}.  

\item Let $\Ext_w$, $\Ext_q$ be strong seeded extractors which use $s$ bits to extract from a $(d, 4s)$ source and outputs $3s$ bits with error $\e'$.

\item Let $\Ext'$ be a strong seeded extractor which uses $r=c \log (3s/\e')$ random bits to extract from an $(3s, 1.5s)$ source and outputs $s$ bits with error $\e'$, from Theorem~\ref{thm:optext}.  

\item Let $\Ext''$ be a strong seeded extractor which uses $s \geq c \log(d/\e')$ random bits to extract from a $(d, 0.15d)$ source and outputs $0.1d$ bits with error $\e'$.

\item Let $\bip$ be the two source extractor from Theorem~\ref{thm:ip}, set up to extract from two $0.3d$-bit sources and output $0.05d$ bits. 

\end{itemize}

\begin{enumerate}

\item Let $\ell=\log a$.\footnote{Without loss of generality we assume that $a$ is a power of $2$. Otherwise add $0$ to the string until the length is a power of $2$.} Let $X_1$ be a slice of $X$ with length $0.3d$, and $Y_1$ be a slice of $Y$ with length $0.3d$. Compute $Z=\bip(X_1, Y_1)$.

Using $Z, Y$ as $Q, W$ (and $S_1$ is a small slice of $Q$) and $\Ext_w, \Ext_q$ as the extractors, run alternating extraction between $Z$ and $Y$ for $2\ell+1$ steps, and output $R_0, R_1, R_2, \cdots, R_{2\ell}=\laext(Y, Z)$, where each $R_i$ has $3s$ bits. Similarly, using $Z, X$ as $Q, W$ (and $S_1$ is a small slice of $Q$) and $\Ext_w, \Ext_q$ as the extractors, run alternating extraction between $Z$ and $X$ for $\ell+1$ steps, and output $S_0, S_1, S_2, \cdots, S_{\ell}=\laext(X, Z)$, where each $S_i$ has $3s$ bits.

\item Use $S_0, R_0, \alpha$ to obtain an $a \times s$ matrix $V^0$, where for any $i \in [a]$, $V^0_i=\flip(S_0, R_0, \alpha_i)$ and outputs $s$ bits.

\item For $j=1, \ldots, \ell$ do the following. Merge the matrix $V^{j-1}$ two rows by two rows: Note that $V^{j-1}$ has $a/2^{j-1}$ rows, for $i=1, \ldots, a/2^j$, compute $\overline{V}^{j-1}_i=\nipm(V^{j-1}_{2i-1}, V^{j-1}_{2i}, R_{2j-1})$ which outputs $r$ bits, and $\tilde{V}^{j-1}_i=\Ext(R_{2j}, V'^{j-1}_i)$ which has $r$ bits. Finally compute $V^j_i=\Ext'(S_{j}, \tilde{V}^{j-1}_i)$ which has $s$ bits.

\item Compute $\hat{V}=\Ext''(X, \Ext_w(Y, V^{\ell}))$.
\end{enumerate}

We now have the following lemma.

\BL \label{lem:advcb}
There exists a constant $C>1$ such that for any $0< \e< 1/2$ and any $a, d \in \N$ such that $d \geq C\log a \log(d a/\e)$, there is an explicit construction of a function $\acb: \bits^d \times \bits^d \times \bits^a \to \bits^{d/10}$ that satisfies the following. Let $Y, Y'$ be $d$-bit random variables such that $H_{\infty}(Y) \geq 0.9d$, and $X, X'$ be $d$-bit random variables with $H_{\infty}(X) \geq 0.9d$. Assume that $(X, X')$ is independent of $(Y, Y')$. Then, for any pair of distinct $a$-bit strings $\alpha, \alpha'$,

\[(\acb(X,Y,\alpha), \acb(X',Y',\alpha'), Y, Y') \approx_{\eps} (U,\acb(X',Y',\alpha'), Y, Y').\] 
\EL

\begin{proof}
We show that with appropriately chosen parameters $s, \e'$ the above construction gives the desired correlation breaker with advice. We will use letters with prime to denote all the corresponding random variables produced by running the same algorithm on $(X', Y')$ instead of $(X, Y)$. Note that both $X_1$ and $Y_1$ has min-entropy at least $0.2d$. Thus by Theorem~\ref{thm:ip} we have that 

\[(Z, X_1)-(U, X_1) \leq 2^{-\Omega(d)} \text{ and } (Z, Y_1)-(U, Y_1) \leq 2^{-\Omega(d)}.\]

We now fix $(Y_1, Y'_1)$, and conditioned on this fixing $(Z, Z')$ is a deterministic function of $(X_1, X'_1)$, thus independent of $(Y, Y')$. Moreover, $Z$ is close to uniform and the average conditional min-entropy of $Y$ is at least $0.9d-2 \times 0.3d=0.3d$. 

Now by Lemma~\ref{altext}, as long as $0.3d \geq 4s+ 2 (2\ell+1) 3s+2 \log(\frac{1}{\e'})$ and $0.05d \geq 4s+ 2 (2\ell+1) 3s+2 \log(\frac{1}{\e'})$, we have that  for any $0 \leq j \leq 2\ell-1$,

\[(Z, Z',  \{R_0, R_0', \cdots, R_{j}, R'_j\}, R_{j+1}) \approx_{O(\ell \e')} (Z, Z',  \{R_{0}, R'_0, \cdots, R_{j}, R'_j\}, U_s).\]

By a hybrid argument and  the triangle inequality, we have that

\[(Z, Z',  R_0, R_0', \cdots, R_{2\ell}, R'_{2\ell}) \approx_{O(\ell^2 \e')} (Z, Z',  U_s, R'_0, \cdots, U_s, R'_{2\ell}),\] where each $U_s$ is independent of all the previous random variables (but may depend on later random variables). From now on, we will proceed as if each $R_{j+1}$ is uniform given $(Z, Z',  \{R_0, R_0', \cdots, R_{j}, R'_j\})$, since this only adds $O(\ell^2 \e')$ to the final error.

Note that conditioned on the fixing of $(Z, Z')$, we have that $\{(R_i, R_i'), i=0, \ldots, 2\ell\}$ is a deterministic function of $(Y, Y')$, thus independent of $(X, X')$. 

By symmetry, we can repeat the above argument while switching the role of $X$ and $Y$. Specifically, we can fix $(X_1, X'_1)$, and conditioned on this fixing $(Z, Z')$ is a deterministic function of $(Y_1, Y'_1)$, thus independent of $(X, X')$. Moreover, $Z$ is close to uniform and the average conditional min-entropy of $X$ is at least $0.9d-2 \times 0.3d=0.3d$. 

Now again by Lemma~\ref{altext}, as long as $0.3d \geq 4s+ 2 (\ell+1) 3s+2 \log(\frac{1}{\e'})$ and $0.05d \geq 4s+ 2 (\ell+1) 3s+2 \log(\frac{1}{\e'})$, we have that  for any $0 \leq j \leq \ell-1$,

\[(Z, Z',  \{S_0, S_0', \cdots, S_{j}, S'_j\}, S_{j+1}) \approx_{O(\ell \e')} (Z, Z',  \{S_{0}, S'_0, \cdots, S_{j}, S'_j\}, U_s).\]

By a hybrid argument and  the triangle inequality, we have that

\[(Z, Z',  S_0, S_0', \cdots, S_{\ell}, S'_{\ell}) \approx_{O(\ell^2 \e')} (Z, Z',  U_s, S'_0, \cdots, U_s, S'_{\ell}),\] where each $U_s$ is independent of all the previous random variables (but may depend on later random variables). From now on, we will proceed as if each $S_{j+1}$ is uniform given $(Z, Z',  \{S_0, S_0', \cdots, S_{j}, S'_j\})$, since this only adds $O(\ell^2 \e')$ to the final error.

Note that now conditioned on the fixing of $(Z, Z')$, we have that $\{(S_i, S_i'), i=0, \ldots, \ell\}$ is a deterministic function of $(X, X')$, thus independent of $(Y, Y')$. Therefore, we can conclude that conditioned on the fixing of $(X_1, X'_1, Y_1, Y'_1, Z, Z')$, we have that $\{(R_i, R_i'), i=0, \ldots, 2\ell\}$ is a deterministic function of $(Y, Y')$, and $\{(S_i, S_i'), i=0, \ldots, \ell\}$ is a deterministic function of $(X, X')$, thus they are independent. Moreover each $R_i$ and $S_i$ is close to uniform given the previous random variables.

We now have the following claim.

\BCM
For all $i \in [a]$ we have that 
\[\left |V^0_i-U_s \right | \le \e'.\]
Furthermore, there exists an $i \in [a]$ such that 
\[\left |(V^0_i,V'^0_i,R_0,R'_0) - (U_s,V'^0_i,R_0,R'_0)\right | \le \e'.\]
\ECM

Indeed, since $\alpha \neq \alpha'$ there exists an $i \in [a]$ such that $\alpha_i \neq \alpha'_i$. Thus by Theorem~\ref{flip}, and noticing that $3s \geq C_{\ref{flip}}\log(3s/\e')$, the claim follows.  Furthermore, notice that now conditioned on the fixing of $(R_0,R'_0)$, $(V^0,V'^0)$ is a deterministic function of $(S_0, S'_0)$, and thus independent of $\{(R_i, R_i'), i=1, \ldots, 2\ell\}$. We now have the following claim.

\BCM \label{clm:error}
Assume that for some $j \leq \ell$, we have that for all $i$, 
\[\left |(V^j_i,\{R_0,R'_0, \cdots, R_{2j}, R'_{2j}\}) - (U_s,\{R_0,R'_0, \cdots, R_{2j}, R'_{2j}) \right | \le \e_j.\]
Furthermore there exists an $i$ such that 
\[\left |(V^j_i,V'^j_i,\{R_0,R'_0, \cdots, R_{2j}, R'_{2j}\}) - (U_s,V'^j_i,\{R_0,R'_0, \cdots, R_{2j}, R'_{2j}) \right | \le \e_j.\]
Then for all $i$, we have that
\[\left |(V^{j+1}_i,\{R_0,R'_0, \cdots, R_{2(j+1)}, R'_{2(j+1)}\}) - (U_s,\{R_0,R'_0, \cdots, R_{2(j+1)}, R'_{2(j+1)}\}) \right | \le 2(\e_j+2\e').\]
Furthermore there exits an $i$ such that
\[\left |(V^{j+1}_i,V'^{j+1}_i,\{R_0,R'_0, \cdots, R_{2(j+1)}, R'_{2(j+1)}\}) - (U_s,V'^{j+1}_i,\{R_0,R'_0, \cdots, R_{2(j+1)}, R'_{2(j+1)}\}) \right | \le 2(\e_j+2\e').\]
\ECM

To see the claim, we focus on the index $i$ where the corresponding row $V^j_i$ is close to uniform given $V'^j_i$. The properties of the other rows can be obtained using similar and simpler arguments. Notice that conditioned on the fixing of $\{R_0,R'_0, \cdots, R_{2j}, R'_{2j}\}$, we have that $(V^j,V'^j)$ is a deterministic function of $(S_0, S'_0, \cdots, S_j, S'_j)$, and thus independent of $(R_{2j+1}, R'_{2j+1})$. Furthermore, by the property of the look-ahead extractor, we know that $R_{2j+1}$ is uniform. Now by Theorem~\ref{thm:nipm}, and noticing that $s \geq 8c\log(3s/\e')$, we know that whenever the $\nipm$ merges the two rows in which one row of $V^j$ is uniform given the corresponding row of $V'^j$, the output obtained from $V^j$ will be uniform given the output obtained from $V'^j$. Thus, there exists an $i$ such that 

\[\left | (\overline{V}^j_i, \overline{V'}^j_i, R_{2j+1}, R'_{2j+1}) -(U_{r}, \overline{V'}^j_i, R_{2j+1}, R'_{2j+1}) \right | \le 2\e_j+2 \e'.\]

Now we fix $(R_{2j+1}, R'_{2j+1})$, and conditioned on this fixing $(\overline{V}^j, \overline{V'}^j)$ is a deterministic function of $(S_0, S'_0, \cdots, S_j, S'_j)$, and thus independent of $(R_{2(j+1)}, R'_{2(j+1)})$. Moreover now again by the property of the look-ahead extractor, we know that $R_{2(j+1)}$ is uniform. Therefore, we can first fix $ \overline{V'}^j_i$ and then $\tilde{V'}^{j}_i=\Ext(R'_{2(j+1)}, V'^{j}_i)$. Conditioned on this fixing we have that $\overline{V}^j_i$ is still uniform, and that $R_{2(j+1)}$ has average conditional min-entropy at least $3s-r=3s-c \log (3s/\e') \geq 23c \log (2s/\e')$. Therefore, by Theorem~\ref{thm:optext} we have that 

\[\left |(\tilde{V}^j_i, \tilde{V'}^j_i, \overline{V}^j_i, \overline{V'}^j_i)-(U_{r}, \tilde{V'}^j_i, \overline{V}^j_i, \overline{V'}^j_i) \right | \leq \e'.\]

Now we can fix $(\overline{V}^j_i, \overline{V'}^j_i)$ and conditioned on this fixing, $(\tilde{V}^j_i, \tilde{V'}^j_i)$ is a deterministic function of $(R_{2(j+1)}, R'_{2(j+1)})$, and thus independent of $(S_{j+1}, S'_{j+1})$. Thus we can first fix $ \tilde{V'}^j_i$ and then $V'^{j+1}_i=\Ext'(S'_{j+1}, \tilde{V'}^{j}_i)$. Note that after this fixing $\tilde{V}^j_i$ is still close to uniform, moreover the average conditional min-entropy of $S_{j+1}$ is at least $3s-s=2s$. Thus by Theorem~\ref{thm:optext} we have that

\[\left |(V^{j+1}_i, V'^{j+1}_i, \tilde{V}^j_i, \tilde{V'}^j_i)-(U_{s}, V'^{j+1}_i, \tilde{V}^j_i, \tilde{V'}^j_i) \right | \leq \e'.\]

Note that conditioned on the fixing of $(\tilde{V}^j_i, \tilde{V'}^j_i)$, we have that $(V^{j+1}_i, V'^{j+1}_i)$ is a deterministic function of $(S_{j+1}, S'_{j+1})$, and thus independent of $(R_{2(j+1)}, R'_{2(j+1)})$. Since we have fixed all the $\{R_0,R'_0, \cdots, R_{2j}, R'_{2j}\}$ before, by adding all the errors we obtain that 

\[\left |(V^{j+1}_i,V'^{j+1}_i,\{R_0,R'_0, \cdots, R_{2(j+1)}, R'_{2(j+1)}\}) - (U_s,V'^{j+1}_i,\{R_0,R'_0, \cdots, R_{2(j+1)}, R'_{2(j+1)}\}) \right | \le 2(\e_j+2\e').\]

Now note that by the end of the iteration of step 3, $V^{\ell}$ has only one row. From Claim~\ref{clm:error} we see that (by solving the recursion of the errors) 

\[\left |(V^\ell,V'^\ell,\{R_0,R'_0, \cdots, R_{2\ell}, R'_{2\ell}\}) - (U_s,V'^\ell,\{R_0,R'_0, \cdots, R_{2\ell}, R'_{2\ell}\}) \right | \le 10a \e'.\]

Note that conditioned on the fixing of $X_1, Y_1, X_1', Y_1', \{R_0,R'_0, \cdots, R_{2\ell}, R'_{2\ell}\}$, we have that $(V^\ell,V'^\ell)$ is a deterministic function of $(X, X')$, and thus independent of $(Y, Y')$. Furthermore the average conditional min-entropy of $Y$ is at least $0.9d-2 \times 0.3d-2(2\ell+1)3s=0.3d-(12\ell+6)s$. Thus we can first fix $V'^\ell$ and then $\Ext_w(Y, V'^\ell)$, and conditioned on this fixing we have that $V^\ell$ is still close to uniform and independent of $Y$, and the average conditional min-entropy of $Y$ is at least $0.3d-(12\ell+9)s$. Now as long as $0.3d-(12\ell+9)s \geq 4s$, by Theorem~\ref{thm:optext} we have that

\[\left |\Ext_w(Y, V^\ell), \Ext_w(Y', V'^\ell), V^\ell,V'^\ell) - (U_{3s},\Ext_w(Y', V'^\ell), V^\ell,V'^\ell\}) \right | \le \e'.\]

Finally, notice that conditioned on the further fixing of $V^\ell,V'^\ell$, we have that $(\Ext_w(Y, V^\ell), \Ext_w(Y, V'^\ell))$ is a deterministic function of $(Y, Y')$, and thus independent of $(X, X')$.  Furthermore the average conditional min-entropy of $X$ is at least $0.9d-2 \times 0.3d-2(\ell+1)3s=0.3d-(6\ell+6)s$. Thus we can first fix $\Ext_w(Y', V'^\ell)l$ and then $\hat{V'}=\Ext''(X', \Ext_w(Y, V'^\ell))$, and conditioned on this fixing we have that $\Ext_w(Y, V^\ell)$ is still close to uniform and independent of $X$, and the average conditional min-entropy of $X$ is at least $0.3d-(6\ell+6)s-0.1d=0.2d-(6\ell+6)s$. Thus as long as $0.2d-(6\ell+6)s \geq 0.15d$, Theorem~\ref{thm:optext} we have that

\[\left |\hat{V}, \hat{V'}, \Ext_w(Y, V^\ell), \Ext_w(Y', V'^\ell)) - (U_{0.1d},\hat{V'}, \Ext_w(Y, V^\ell), \Ext_w(Y', V'^\ell)) \right | \le \e'.\]

Note that now conditioned on the fixing of $(\Ext_w(Y, V^\ell), \Ext_w(Y', V'^\ell))$, we have that $(\hat{V}, \hat{V'})$ is a deterministic function of $(X, X')$, and thus independent of $(Y, Y')$. Therefore by adding back all the errors we obtain 

\[\left |\hat{V}, \hat{V'}, Y, Y') - (U_{0.1d},\hat{V'}, Y, Y') \right | \le \e_1,\]
where $\e_1=(10a+2)\e'+O(\ell^2 \e')+2^{-\Omega(d)}$.

Next, in order for all the entropy requirement to hold, we need the following conditions.
\[s \geq \max\{c \log(d/\e'), 8c \log (3s/\e')\}, \text{ and } 0.05d \geq 4s+ 2 (2\ell+1) 3s+2 \log(\frac{1}{\e'})\]
\[0.3d-(12\ell+9)s \geq 4s, \text{ and } 0.2d-(6\ell+6)s \geq 0.15d.\]

The above conditions are satisfied if the following conditions are satisfied.

\[d \geq 240(\ell+1)s, \text{ and } s \geq 8c \log (d/\e').\]

Under this condition, we see that $2^{-\Omega(d)} \leq \e'$, and since $\ell=\log a$ we have that $\ell^2=O(a)$. Thus the total error is $\e_1=O(a)\e'$. Therefore, to make $\e_1=\e$, we can set $\e'=\e/(c' a)$ for some constant $c'>0$. We can now set $s = 9c\log (d/\e')=9c\log(c' d a/\e)$, and the conditions are satisfied as long as $d \geq C \ell \log(d a/\e)=C\log a \log(d a/\e)$ for some constant $C>1$.

\end{proof}

%% file: ext.tex
\section{The Seeded Non-Malleable Extractor}\label{sec:snmext}
In this section we construct our improved seeded non-malleable extractor. First we need the following advice generator from \cite{CGL15}

\begin{theorem}[\cite{CGL15}]\label{adv_gen} There exist a constant $c>0$ such that  for all $n>0$ and any $\epsilon>0$, there exists an explicit function $\adg:\zo^n \times \zo^d \rightarrow \zo^{L}$ with $L=c \log (n/\epsilon)$ satisfying the following: Let $X$ be an $(n,k)$-source, and $Y$ be an independent uniform seed on $d$ bits. Let $Y'$ be a random variable on $d$ bits s.t $Y' \neq Y$, and $(Y, Y')$ is independent of $X$. Then with probability at least $1-\epsilon$, $\adg(X,Y) \neq \adg(X,Y')$. Moreover, there is a deterministic function $g$ such that $\adg(X, Y)$ is computed as follows. Let $Y_1$ be a small slice of $Y$ with length $O(\log (n/\e))$, compute $Z_1=\Ext(X, Y_1)$ where $\Ext$ is an optimal seeded extractor from Theorem~\ref{thm:optext} which outputs $O(\log (n/\e))$ bits. Finally compute $Y_2=g(Y, Z_1)$ which outputs $O(\log(1/\e))$ bits and let $\adg(X, Y)=(Y_1, Y_2)$.
\end{theorem}

The construction of the non-malleable extractor is as follows.

\begin{itemize}
\item Let $\e'=\e/10$. Assume $k \geq 6d$.

\item Let $\Ext$ be a strong seeded extractor from Theorem~\ref{thm:optext}, which uses $O(\log(n/\e'))$ bits to extract from an $(n, k/3)$ source and outputs $k/4$ bits with error $\e'$.

\item Let $\Ext'$ be a strong seeded extractor from Theorem~\ref{thm:optext}, which uses $O(\log(n/\e'))$ bits to extract from an $(n, k)$ source and outputs $d$ bits with error $\e'$.

\item Let $\adg$ be the advice generator from Theorem~\ref{adv_gen}, with error $\e'$.

\item Let $\acb$ be the correlation breaker with advice from Lemma~\ref{lem:advcb}, with error $\e'$.
\end{itemize}

\begin{enumerate}
\item Compute $\adg(X, Y)$ with error $\e'$. Specifically, first compute $X_1=\Ext'(X, Y_1)$, except now it outputs $Z=\Ext(X, Y_1)$ with $d$ bits. Let $Z_1$ be a slice of $Z$ with $O(\log (n/\e'))$ bits and as in Theorem~\ref{adv_gen}, compute $Y_2=g(Y, Z_1)$ which outputs $O(\log(1/\e'))$ bits and let $\adg(X, Y)=(Y_1, Y_2)=\alpha$.

\item Compute $V=\acb(Y, Z, \alpha)$ which outputs $d/10$ bits.

\item Output $W=\Ext(X, V)$ which outputs $k/4$ bits
\end{enumerate}

We now have the following theorem.

\BT \label{thm:snmmain}
There exists a constant $C>1$ such that for any $n, k \in \N$ and $0<\e<1$ with $k \geq C(\log n+\log \log(1/\e) \log(1/\e))$, there is an explicit construction of a strong seeded $(k, \e)$ non-malleable extractor $\zo^n \times \zo^d \to \zo^m$ with $d=C(\log n+\log \log(1/\e) \log(1/\e))$ and $m \geq k/4$.
\ET

\begin{proof}
Again, we use letters with prime to denote random variables produced with $(X, Y')$ instead of $(X, Y)$. First note that by Theorem~\ref{thm:optext}, we have that
\[(Z, Y_1) \approx_{\e'} (U_d, Y_1).\]

We will now proceed as if $Z$ is uniform given $Y_1$, since this only adds error $\e'$. We now fix $(Y_1, Y_1')$. Note that conditioned on this fixing, $(Z, Z')$ is a deterministic function of $X$, and thus independent of $(Y, Y')$. Moreover by Lemma~\ref{lem:condition} with probability $1-\e'$, the min-entropy of $Y$ is at least $d-O(\log (n/\e'))$. Now we fix $(Z_1, Z_1')$, and note that conditioned on this fixing, $(Y_2, Y'_2)$ is a deterministic function of $(Y, Y')$, and thus independent of $(X, Z, Z')$. Moreover again by by Lemma~\ref{lem:condition} with probability $1-\e'$, the min-entropy of $Z$ is at least $d-O(\log (n/\e'))$. Finally we fix $(Y_2, Y_2')$. Note that conditioned on this fixing, $(Y, Y')$ is still independent of $(X, Z, Z')$. Moreover by Lemma~\ref{lem:condition} with probability $1-\e'$, the min-entropy of $Y$ is at least $d-O(\log (n/\e'))$. Also note that by Theorem~\ref{adv_gen}, with probability at least $1-\e'$ over the fixing of $(Y_1, Z_1, Y_2, Y'_1, Z'_1, Y'_2)$, we have that $\alpha=(Y_1, Y_2) \neq (Y_1', Y_2')=\alpha'$. Thus, as long as $d \geq C \log(n/\e')$ for some constant $C>1$, altogether we can conclude that with probability at least $1-4\e'$, we have that 

\begin{itemize}
\item $\alpha \neq \alpha'$, where $\alpha, \alpha'$ each has $a=c\log(n/\e')$ bits.

\item $X$ is still independent of $(Y, Y')$, and $(Z, Z')$ is a deterministic function of $X$.

\item $H_{\infty}(Y) \geq 0.9d$ and $H_{\infty}(Z) \geq 0.9d$.
\end{itemize}

Thus, as long as $d \geq C'\log a \log(d a/\e')$ where $C'$ is the constant in Lemma~\ref{lem:advcb}, we have that 

\[(V, V', Z, Z') \approx_{\e'} (U,V', Z, Z').\] 

Note that conditioned on the fixing of $(Z, Z')$, we have that $(V, V')$ is a deterministic function of $(Y, Y')$, and thus independent of $X$. Moreover the average conditional min-entropy of $X$ is at least $k-2d \geq 2k/3$. Thus now we can first fix $V'$ and then $W'=\Ext(X, V')$. Note that after this fixing $X$ and $(Y, Y', V)$ are still independent. Moreover $V$ is still close to uniform and the average conditional min-entropy of $X$ is at least $2k/3-k/4> k/3$. Thus by Theorem~\ref{thm:optext} we have that
\[(W, W', V, V') \approx_{\e'} (U, W', V, V').\]

Note that conditioned on the fixing of $(V, V')$, we have that $(W, W')$ is a deterministic function of $X$, thus independent of $(Y, Y')$. Therefore by adding back all the errors we get that 
\[(W, W', Y, Y') \approx_{7\e'} (U, W', Y, Y').\]

Since $\e'=\e/10$ we have that
\[(W, W', Y, Y') \approx_{\e} (U, W', Y, Y').\]

Now let's decide the seed length $d$. We need to have that 
\[d \geq C \log(n/\e') \text{ and } d \geq C'\log a \log(d a/\e'),\] where $a=c\log(n/\e')$ and $\e'=\e/10$. Since our new non-malleable extractor is better than the construction in \cite{CGL15}, which has seed length $d=O(\log^2(n/\e'))$, we can first assume that $d=O(\log^2(n/\e))$ and we will use the inequality $d \geq C'\log a \log(d a/\e')$ to compute the minimum $d$ and verify the condition that $d=O(\log^2(n/\e'))$ does hold.

In this case, we see that $\log(d a/\e')=\log (O(\log^3(n/\e')/\e'))=O(\log \log (n/\e')+\log(1/\e'))$, and $\log a=O(\log \log (n/\e'))$. Thus we need 
\[d \geq C_1((\log \log (n/\e'))^2+\log \log (n/\e')\log(1/\e'))\] for some constant $C_1>1$.

Note that if $\e' < 1/n$, then $\log \log (n/\e')=O(\log \log(1/\e'))$ and thus $(\log \log (n/\e'))^2 < \log \log (n/\e')\log(1/\e')$; and if $\e' \geq 1/n$ then $(\log \log (n/\e'))^2 < \log n$. Thus we have 
\[(\log \log (n/\e'))^2< \log n + \log \log (n/\e')\log(1/\e').\]
Now consider $ \log \log (n/\e')\log(1/\e')$. We have that 
\[ \log \log (n/\e')\log(1/\e') \leq \log(\log n \log (1/\e'))\log(1/\e')=(\log \log n+\log \log(1/\e')) \log(1/\e').\]
Now if $\e'< 2^{-\log n/\log \log n}$, then we have that $\log \log (1/\e')> \log \log n-\log \log \log n>0.5\log \log n$. Thus in this case we have that 
\[\log \log (n/\e')\log(1/\e') \leq 3\log \log(1/\e') \log(1/\e').\]

On the other hand, if $\e' \geq 2^{-\log n/\log \log n}$, then we have that 
\[\log \log n \log(1/\e') \leq \log n.\]
Thus combining the two cases we have that 
\[\log \log (n/\e')\log(1/\e') \leq \log n+3\log \log(1/\e') \log(1/\e').\]
Altogether we have 
\[(\log \log (n/\e'))^2+\log \log (n/\e')\log(1/\e') \leq 3 \log n+6\log \log(1/\e') \log(1/\e').\]
Thus, it suffices to set 
\[d=O(\log n+\log \log(1/\e') \log(1/\e'))=O(\log n+\log \log(1/\e) \log(1/\e)).\]
\end{proof}

\section{Non-Malleable Two-Source Extractor and Non-Malleable Code}\label{sec:nmtext}
Formally, non-malleable codes are defined as follows. 

\BD \cite{ADKO15}
Let $\snm_k$ denote the set of trivial manipulation functions on $k$-bit strings, which consists of the identity function $I(x)=x$ and all constant functions $f_c(x)=c$, where $c \in \bits^k$. Let $E: \bits^k \to \bits^m$ be an efficient randomized \emph{encoding} function, and $D: \bits^m \to \bits^k$ be an efficient deterministic \emph{decoding} function. Let ${\cal F}: \bits^m \to \bits^m$ be some class of functions. We say that the pair $(E, D)$ defines an $({\cal F}, k, \e)$-\emph{non-malleable code}, if for all $f \in {\cal F}$ there exists a probability distribution $G$ over $\snm_k$, such that for all $x \in \bits^k$, we have

\[\left |D(f(E(x)))-G(x) \right | \leq \e.\]  
\ED

\begin{remark}
The above definition is slightly different form the original definition in \cite{DPW10}. However, \cite{ADKO15} shows that the two definitions are equivalent.
\end{remark}

We will mainly be focusing on the following family of tampering functions in this paper.

\BD Given any $\ell>1$, let ${\cal S}^\ell_n$ denote the tampering family in the $\ell$-\emph{split-state-model}, where the adversary applies $\ell$ arbitrarily correlated functions $h_1, \cdots, h_{\ell}$ to $\ell$ separate, $n$-bit parts of string. Each $h_i$ can only be applied to the $i$-th part individually.
\ED

Note that although the functions $h_1, \cdots, h_{\ell}$ can be correlated, their correlation does not depend on the original codewords. Thus, they are a convex combination of independent functions, applied to each part of the codeword. Thus, without loss of generality, hereafter we may assume that each $h_i$ is an independent function acting on the $i$-th part of the codeword individually. In this paper we will mainly consider the case of $\ell=2$, i.e., the two-split-state model.

The following theorem was proved by Cheraghchi and Gursuswami \cite{CG14b}, which establishes a connection between seedless non-malleable extractors and non-malleable codes.

\BT \label{connection} Let $\nmExt:\{0,1\}^{n} \times \{ 0,1\}^n \rightarrow \{0,1\}^{m}$  be a polynomial time computable seedless $2$-non-malleable extractor  at min-entropy $n$ with error $\epsilon$. Then there exists an explicit non-malleable code with an efficient decoder in the $2$-split-state model with block length $=2n$, rate  $= \frac{m}{2n}$ and error $=2^{m+1}\epsilon$.
\ET

Using the non-malleable extractor, the non-malleable code in the $2$-split-state model is constructed as follows: For any message $s \in \{ 0,1\}^m$, the encoder $\Enc(s)$ outputs a uniformly random string from the set $\nmExt^{-1}(s) \subset \{ 0,1\}^{2n}$. For any codeword $c \in \{0,1\}^{2n}$, the decoder $\Dec$ outputs $\nmExt(c)$. Thus, for the encoder to be efficient we need to be able to efficiently uniformly sample from the pre-image of any output of the extractor. We will now first describe our construction of the non-malleable extractor and then show how to efficiently uniformly sample from the pre-image.

\subsection{The construction and the analysis of the extractor}
We have the following construction of a non-malleable two-source extractor for two $(n, (1-\gamma)n)$ sources, where $0<\gamma<1$ is some constant. First we need the following construction of an ``invertible" linear seeded extractor.

\BT \label{thm:iext}
There exists a constant $0<\alpha<1$ such that for any $n \in \N$ and $2^{-\alpha n}< \e<1 $ there exists a linear seeded strong extractor $\iext: \bits^n \times \bits^d \to \bits^{0.3 d}$ with $d=O(\log(n/\e))$ and the following property. If $X$ is a $(n,0.9n)$ source and $R$ is an independent uniform seed on $\{ 0,1\}^{d}$, then $$ |(\iext(X,R),R) - (U_{0.3 d},R)| \leq \e.$$ 
Furthermore for any $s \in \{ 0,1\}^{0.3 d}$ and any $r \in  \{ 0,1\}^{d}$, $| \iext(\cdot,r)^{-1}(s)|= 2^{n-0.3 d}$.
\ET

To prove the theorem we need the following definitions and theorems.

\BD[Averaging sampler \cite{Vadhan04}]\label{def:samp} A function $\samp: \{0,1\}^{r} \rightarrow [n]^{t}$ is a $(\mu,\theta,\gamma)$ averaging sampler if for every function $f:[n] \rightarrow [0,1]$ with average value $\frac{1}{n}\sum_{i}f(i) \ge \mu$, it holds that 
$$ \Pr_{i_1,\ldots,i_t \leftarrow \samp(U_{R})}\left [  \frac{1}{t}\sum_{i}f(i) < \mu - \theta \right ] \leq \gamma.$$
$\samp$ has distinct samples if for every $x \in \{ 0,1\}^{r}$, the samples produced by $\samp(x)$ are all distinct.
\ED

\BT [\cite{Vadhan04}] \label{thm:samp} Let $1 \geq \delta \geq 3\tau > 0$. Suppose that $\samp: \zo^r \to [n]^t$ is an $(\mu,\theta,\gamma)$ averaging sampler with distinct samples for $\mu=(\delta-2\tau)/\log(1/\tau)$ and $\theta=\tau/\log(1/\tau)$. Then for every $\delta n$-source $X$ on $\zo^n$, the random variable $(U_r,  X_{Samp(U_r)})$ is $(\gamma+2^{-\Omega(\tau n)})$-close to $(U_r, W)$ where for every $a \in \zo^r$, the random variable $W|_{U_r=a}$ is $(\delta-3\tau)t$-source.
\ET

\BT[\cite{Vadhan04}]\label{thm:sampler} For every $0< \theta< \mu<1$, $\gamma>0$, and $n \in \N$, there is an explicit $(\mu,\theta,\gamma)$ averaging sampler $\samp: \zo^r \to [n]^t$
 that uses
 \begin{itemize}
 \item $t$ distinct samples for any $t \in [t_0, n]$, where $t_0=O(\frac{1}{\theta^2} \log(1/\gamma))$, and
 \item $r=\log (n/t)+\log(1/\gamma)\poly(1/\theta)$ random bits.
 \end{itemize}
\ET

We can now prove Theorem~\ref{thm:iext}.
\begin{proof}[Proof of Theorem~\ref{thm:iext}]
Given the source $X$ and the seed $R$, we construct the extractor $\iext$ as follows. Set $\delta=0.9$ and $\tau=0.1$. Set $\mu=(\delta-2\tau)/\log(1/\tau)$, $\theta=\tau/\log(1/\tau)$ and $\gamma=\e/4$. Now by Theorem~\ref{thm:sampler} there is an explicit $(\mu,\theta,\gamma)$ averaging sampler $\samp: \zo^r \to [n]^t$ that uses $t$ distinct samples for any $t \in [t_0, n]$, where $t_0=O(\frac{1}{\theta^2} \log(1/\gamma))=O(\log(1/\e))$ and $r=\log (n/t)+\log(1/\gamma)\poly(1/\theta)=\log n+O(\log(1/\e))$. We will set $t=0.9 d+1$ and $r=0.1d$. Note that by setting the hidden constant in $d=O(\log(n/\e))$ to be big enough and $\alpha$ to be small enough we can ensure that $0.9d+1 \in [t_0, n]$, and $r \leq 0.1d$. Thus such a sampler can indeed be constructed. 

We now take a slice of $0.1d$ bits from $R$ and let $R=(R_1, R_2)$, where $R_2$ has $0.9 d$ bits. We use $R_1$ to sample $t=0.9d+1$ distinct bits from $X$, and let the sampled bits be $X'$. By Theorem~\ref{thm:samp} we know that $(R_1, X')$ is $\e/4+2^{-\Omega(n)}$-close to $(R_1, W)$ where conditioned on any fixing of $R_1$, $W$ is a $0.6t \geq 0.5 d$ source. We will now proceed as if $(R_1, X')$ is $(R_1, W)$, since this only adds error $\e/4+2^{-\Omega(n)}$.

Next we fix $R_1$, and note that conditioned on this fixing, $X'$ is a deterministic function of $X$, and thus independent of $R_2$. Further $X'$ has entropy rate $0.6$. We now take $R_2$ and let $R_2'$ be $R_2$ padding with a $1$ at the end, thus $R_2'$ also has $t=0.9 d+1$ bits and has min-entropy $0.9 d$. Finally we compute the output $\iext(X, R)$ to be the last $0.3 d$ bits of $R_2' \cdot X'$, where the operation is in the field $\F_{2^t}$.  By the leftover hash lemma we know that 

\[|(\iext(X, R), R_2) -(U, R_2)| \leq 2 \cdot 2^{-0.1 d}.\]
 
Since $r \leq 0.1d$ we have that $2^{-0.1d}< \gamma=\e/4$. Since conditioned on the fixing of $R_2$ we have that $\iext(X, R)$ is a deterministic function of $X$, by adding back all the errors we get 

\[|(\iext(X, R), R) -(U, R)| \leq \e/4+\e/4+2^{-\Omega(n)}.\]

By setting $\alpha$ to be small enough we can ensure the total error is at most $\e$, thus we have

\[|(\iext(X, R), R) -(U, R)| \leq \e.\]

Moreover, for any fixing of the seed $R=r$, the function $\iext(X, r)$ is a linear function in $X$ because it first selects $t$ bits from $X$ and then performs the operation $R_2' \cdot X'$, which is a linear function since the field is $\F_{2^t}$. Finally, the pre-image size for any fixed seed is the same since first, the pre-image size of $X'$ is always $2^{t-0.3 d}$ because $R_2'$ is a fixed non-zero field element, and then given $X'$ to get $X$ it is enough to put any bits for the un-sampled part of $X$. 
\end{proof}

We now have the following construction. Let $(X, Y)$ be two independent $(n, (1-\gamma)n)$ source. 
\begin{itemize}
\item Let $0<\gamma<\alpha<\beta<1/70$ be two constants to be chosen later.

\item Let $\bip$ be the inner product two-source extractor from Theorem~\ref{thm:ip}.

\item Let $\acb$ be the correlation breaker with advice from Lemma~\ref{lem:advcb} with error $\e=2^{-\Omega(n/\log n)}$.

\item Let $\iext$ be the invertible linear seeded extractor form Theorem~\ref{thm:iext}.
\end{itemize}

\begin{enumerate}
\item Let $n_1=\alpha n$. Divide $X$ into $X=(X_1, X_2)$ such that $X_1$ has $n_1$ bits and $X_2$ has $n_2=(1-\alpha)n$ bits. Similarly divide $Y$ into $Y=(Y_1, Y_2)$ such that $Y_1$ has $n_1$ bits and $Y_2$ has $n_2=(1-\alpha)n$ bits.

\item Compute $Z=\bip(X_1, Y_1)$ which outputs $r=\Omega(n) \leq \alpha n/2$ bits.

\item Let $\F$ be the finite field $\F_{2^{\log n}}$. Let $n_0 = \frac{n_2}{\log n}$. Let $\RS: \F^{n_0} \rightarrow \F^{n}$ be the Reed-Solomon code encoding $n_0$ symbols of $\F$ to $n$ symbols in  $\F$ (we  slightly abuse the use of $\RS$ to denote both the code and the encoder). Thus $\RS$ is a $[n,n_0,n-n_0+1]_{n}$ error correcting code. Let $X'_2$ be $X_2$ written backwards, and similarly $Y'_2$ be $Y_2$ written backwards. Let $\overline{X}_2=\RS(X'_2)$ and $\overline{Y}_2=\RS(Y'_2)$.

\item Use $Z$ to sample $r/\log n$ distinct symbols from $\overline{X}_2$ (i.e., use each $\log n$ bits to sample a symbol), and write the symbols as a binary string $\tilde{X}_2$. Note that $\tilde{X}_2$ has $r$ bits. Similarly, use $Z$ to sample $r/\log n$ distinct symbols from $\overline{Y}_2$ and obtain a binary string $\tilde{Y}_2$ with $r$ bits.

\item Let $\tilde{\alpha}=X_1 \circ Y_1 \circ \tilde{X}_2 \circ \tilde{Y}_2$. Divide $X_2$ into $X_2=(X_3, X_4, X_5)$ such that $X_3$ has $n_3=\beta n$ bits, $X_4$ has $n_4=30\beta n$ bits and $X_5$ has $n_5=(1-\alpha-31\beta)n$ bits. Similarly divide $Y_2=(Y_3, Y_4, Y_5)$ such that $Y_3$ has $n_3$ bits, $Y_4$ has $n_4$ bits and $Y_5$ has $n_5$ bits.

\item Compute $V=\acb(X_3, Y_3, \tilde{\alpha})$ which outputs $d=n_3/10=\beta n/10$ bits.

\item Finally compute $W=\iext(Y_4, V)$ which outputs $\Omega(d) < d/2$ bits.
\end{enumerate}

We now have the following theorem.

\BT \label{thm:nmtext}
There exists a constant $0< \gamma< 1$ and a non-malleable two-source extractor for $(n, (1-\gamma)n)$ sources with error $2^{-\Omega(n/\log n)}$ and output length $\Omega(n)$.
\ET

\begin{proof}
We show that the above construction is such a non-malleable two-source extractor. As usual, we will use letters with prime to denote random variables produced from $(X', Y')$. Without loss of generality we assume that $X \neq X'$. The case where $Y \neq Y'$ can be handled in the same way by symmetry.

First we argue that with probability $1-2^{-\Omega(n/\log n)}$ over $\tilde{\alpha}, \tilde{\alpha'}$, we have that $\tilde{\alpha} \neq \tilde{\alpha'}$. To see this, note that if $X_1 \neq X_1'$ or $Y_1 \neq Y_1'$ then $\tilde{\alpha} \neq \tilde{\alpha'}$. Otherwise, since $X \neq X'$ we must have $X_2 \neq X_2'$. Thus by the property of the $\RS$ code we know that $\overline{X}_2$ and $\overline{Y}_2$ must differ in at least $n-n_0>0.9 n$ symbols. Also, since $X_1=X_1'$ and $Y_1=Y_1'$ we have $Z=Z'$. Now if $\alpha \geq 3\gamma$ then both $X_1$ and $Y_1$ has min-entropy rate at least $2/3$, thus by Theorem~\ref{thm:ip} we know that

\[(Z, X_1) \approx_{2^{-\Omega(n)}} (U_r, X_1).\]

We can now fix $X_1$, and conditioned on this fixing $Z$ is a deterministic function of $Y$, thus independent of $X_2$. Therefore now we can use $Z$ to sample from $\overline{X}_2$. If $Z$ is uniform then by a Chernoff bound we know that 

\[\Pr[\tilde{X}_2 \neq \tilde{X'}_2] \geq 1-2^{-r/\log n}=1-2^{-\Omega(n/\log n)}.\]

Thus the total probability that $\tilde{\alpha} \neq \tilde{\alpha'}$ is at least $1-2^{-\Omega(n/\log n)}-2^{-\Omega(n)}=1-2^{-\Omega(n/\log n)}$.

Moreover, by choosing $\alpha< \beta/50$, we can ensure that $r \leq \alpha n < \beta n/50$. Now by Lemma~\ref{lem:condition} we know that conditioned on the fixing of $(\tilde{\alpha}, \tilde{\alpha'})$, with probability $1-2^{-\Omega(n)}$, we have that $H_{\infty}(X_3) \geq \beta n-\gamma n-\alpha n-3r \geq 0.9 \beta n$ and similarly $H_{\infty}(Y_3) \geq 0.9 \beta n$. Moreover $(X, X')$ and $(Y, Y')$ are still independent.

Now we will use Lemma~\ref{lem:advcb}. Note that the length of the advice string is $a=2\alpha n+2r \leq 3\alpha n$, and $X_3, Y_3$ each has $\beta n$ bits. Thus by choosing the error $\e=2^{-\Omega(n/\log n)}$ appropriately we can ensure that 

\[\beta n \geq C\log a \log(\beta n a/\e),\]
where $C$ is the constant in  Lemma~\ref{lem:advcb}. When this condition holds, by Lemma~\ref{lem:advcb} we have that

\[(V, V', Y_3, Y_3') \approx_{\e} (U_d, V', Y_3, Y_3').\]

We now fix $(Y_3, Y_3')$. Note that conditioned on this fixing, $(V, V')$ is a deterministic function of $(X, X')$, and thus independent of $(Y, Y')$. Moreover the average conditional min-entropy of $Y_4$ is at least $n_4-\gamma n-\alpha n-2r-\beta n \geq n_4 -2\alpha n-\beta n$. Note that $n_4=30 \beta n$. Thus by choosing $\alpha< \beta/50$ we can ensure that (by Lemma~\ref{lem:condition}) with probability $1-2^{-\Omega(n)}$, $Y_4$ has min-entropy rate at least $0.95$.

Now we can fix $V'$ and then $W'$. Note that conditioned on this fixing, $V$ is still close to uniform, and independent of $Y_4$. Furthermore since the length of $W'$ is at most $d/2=\beta n/20$, again by Lemma~\ref{lem:condition} we have that with probability $1-2^{-\Omega(n)}$, $Y_4$ has min-entropy rate at least $0.9$. Thus now by Theorem~\ref{thm:iext} we have that

\[(W, V) \approx_{2^{-\Omega(n)}} (U, V).\]

Note that conditioned on the fixing of $V$, $W$ is a deterministic function of $Y$. Since we have already fixed $(V', W')$, by adding back all the errors we get that 

\[(W, W', X, X') \approx_{2^{-\Omega(n/\log n)}} (U, W', X, X').\]
\end{proof}

\subsection{Efficiently sampling algorithm and the non-malleable code}
We now show that given an output of the non-malleable two-source extractor, we can efficiently uniformly sample from the pre-image of this output. First we have the following main lemma.

\BL \label{lem:samp}
Given any arbitrary fixing of $(X_1, \tilde{X}_2, X_3, Y_1, \tilde{Y}_2, Y_3, W)$, there is an efficient procedure to uniformly sample from the pre-image $(X, Y)$. Moreover, for any fixing of $(X_1, \tilde{X}_2, X_3, Y_1, \tilde{Y}_2, Y_3, W)$, the pre-image has the same size.
\EL

\begin{proof}
Assume that we are given $(X_1, \tilde{X}_2, X_3, Y_1, \tilde{Y}_2, Y_3, W)=(x_1, \tilde{x}_2, x_3, y_1, \tilde{y}_2, y_3, w)$ for arbitrary $(x_1, \tilde{x}_2, x_3, y_1, \tilde{y}_2, y_3, w)$. We need to sample from the corresponding $(X_4, X_5, Y_4, Y_5)$. First we can compute $z=\bip(x_1, y_1)$ which tells us what symbols of the $\RS$ codes are sampled. Next, we can compute $v=\acb(x_3, y_3, \tilde{\alpha})$ where $\tilde{\alpha}=x_1 \circ y_1 \circ \tilde{x}_2 \circ \tilde{y}_2$. Now note that $W=\iext(Y_4, V)$, therefore by Theorem~\ref{thm:iext} we can efficiently and uniformly sample the pre-image of $w$, which is $Y_4$, by inverting a system of linear equations. Also, Theorem~\ref{thm:iext} guarantees that for any $(v, w)$ the pre-image has the same size. 

Now once we have sampled $Y_4=y_4$, we will continue to sample $(X_4, X_5, Y_5)$. Since these are different bits in $(X, Y)$ than the bits we have already obtained, they can almost be sampled arbitrarily, except they need to satisfy the linear constraints imposed by the $\RS$ codes: $\tilde{Y}_2=y_2$ and $\tilde{X}_2=x_2$. We first look at the $Y$ part. Note that $\tilde{Y}_2=y_2$ gives us $r/\log n \leq \alpha n/(2 \log n)< n/(4 \log n)$ equations in the field $\F_{2^{\log n}}$. Also note that now $(Y_1, Y_3, Y_4)$ are fixed and $Y_5$ are the variables. Since the length of $Y_5$ is $n_5=n-\alpha n-\beta n-30 \beta n> n/2$ (as $\beta < 1/70$), this gives us at least $n/(2 \log n)$ variables in the field $\F_{2^{\log n}}$. Finally, note that when we encode $Y_2$ using the $\RS$ code, we encode it as $\RS(Y'_2)$ where $Y'_2$ is $Y_2$ written backwards. Thus the coefficient matrix of the equations with variables in $Y_5$ is 
  $$
 G = 
 \begin{pmatrix}
  1 & 1 & \cdots & 1 \\
  \alpha_1 & \alpha_{2} & \cdots & \alpha_{s} \\
  \vdots  & \vdots  & \ddots & \vdots  \\
  \alpha_1^{t} & \alpha_2^{t} & \cdots & \alpha_s^{t}
 \end{pmatrix}
$$ where $s=r/\log n$, $t=n_5/\log n$, and $\alpha_1,\ldots,\alpha_{s}$ are distinct field elements of $\F_{2^{\log n}}$. 

Note that $t=n_5/\log n>n/(2 \log n)>r/\log n=s$, thus all the columns in the matrix are linearly independent, and the kernel of the matrix has dimension exactly $t-s$ for any $\alpha_1,\ldots,\alpha_{s}$. Therefore, we can efficiently sample $Y_5$ by inverting the system of linear equations, and moreover for any fixing of $(Y_1, Y_3, Y_4, Z)=(y_1, y_3, y_4, z)$ the pre-image always has the same size.

The argument for sampling the $X$ part is exactly the same, except now $X$ has more variables ($(X_4, X_5)$) than $Y$.
\end{proof}

We now have the following main theorem.

\BT \label{thm:msample}
Given any output $W=w$ of the non-malleable two-source extractor, there is an efficient procedure to uniformly sample from the pre-image $(X, Y)$.
\ET

\begin{proof}
The sampling procedure is as follows. We first uniformly randomly generate $(X_1, \tilde{X}_2, X_3, Y_1, \tilde{Y}_2, Y_3)=(x_1, \tilde{x}_2, x_3, y_1, \tilde{y}_2, y_3)$, then we use Lemma~\ref{lem:samp} to generate $(X, Y)$. By Lemma~\ref{lem:samp}, for any fixing of $(X_1, \tilde{X}_2, X_3, Y_1, \tilde{Y}_2, Y_3, W)$, the pre-image has the same size. Thus indeed this procedure uniformly samples from the pre-image $(X, Y)$.
\end{proof}

Combining Theorem~\ref{connection}, Theorem~\ref{thm:nmtext}, and Theorem~\ref{thm:msample}, we immediately obtain the following theorem.

\BT \label{thm:nmcode}
For any $n \in \N$ there exists an explicit non-malleable code with efficient encoder/decoder in the $2$-split-state model with block length $2n$, rate  $\Omega(1/\log n)$ and error $=2^{-\Omega(n/\log n)}$.
\ET

\section{$t$-Non-Malleable Extractors and Applications to Independent Source Extractors}\label{sec:indext}
In this section, we extend our results to the case of $t$ tampering functions, and use them to obtain improved results of independent source extractors.

We first prove that any $s$-source non-malleable extractor with sufficiently small error must be a strong $s$-source non-malleable extractor. Formally, we have

\BT \label{thm:nmstrong} Suppose $\nmExt: (\zo^n)^s \to \zo^m$ is an $s$-source non-malleable extractor with error $\e$ for min-entropy $k$. Then for any $k' \geq k$, $\nmExt$ is a strong $s$-source non-malleable extractor for min-entropy $k'$ with error $2^{2m} (\e+2^{k+1-k'})$. 
\ET

\begin{proof}
Let $X_1, \cdots, X_s$ be independent $(n, k')$ sources and $X'_1=f_1(X_1), \cdots, X'_s=f_s(X_s)$ where for each $i$, $f_i : \zo^n \to \zo^n$ is a deterministic function such that at least one of them has no fixed point. Consider any $i$. Let $X_{<i}=(X_1, \cdots, X_{i-1})$, $X_{>i}=(X_{i+1}, \cdots, X_s)$ and similarly $X'_{<i}=(X'_1, \cdots, X'_{i-1})$, $X'_{>i}=(X'_{i+1}, \cdots, X'_s)$. Now for any $(z, z') \in (\zo^m)^2$, define the set of bad $y$'s for $(z, z')$ to be 

\[B_{z, z'}=\{y: \left |\Pr[\nmExt(X_{<i}, y, X_{>i})=z, \nmExt(X'_{<i}, f_i(y), X'_{>i})=z'] -2^{-m}\Pr[\nmExt(X'_{<i}, f_i(y), X'_{>i})=z'] \right |>\e.\}\]

We have the following claim.

\BCM
For any $(z, z')$, we have $|B_{z, z'}| < 2^{k+1}$.
\ECM

Suppose not, then define 

\[B^+_{z, z'}=\{y: \Pr[\nmExt(X_{<i}, y, X_{>i})=z, \nmExt(X'_{<i}, f_i(y), X'_{>i})=z'] -2^{-m}\Pr[\nmExt(X'_{<i}, f_i(y), X'_{>i})=z'] >\e.\}\]

and 

\[B^-_{z, z'}=\{y: \Pr[\nmExt(X_{<i}, y, X_{>i})=z, \nmExt(X'_{<i}, f_i(y), X'_{>i})=z'] -2^{-m}\Pr[\nmExt(X'_{<i}, f_i(y), X'_{>i})=z'] < -\e.\}\]

We have that either $|B^+_{z, z'}| \geq 2^k$ or $|B^-_{z, z'}| \geq 2^k$. Without loss of generality assume that $|B^+_{z, z'}| \geq 2^k$. Then, let $Y$ be the uniform distribution over $B^+_{z, z'}$. We have that $Y$ is an $(n, k)$ source, but 

\begin{align*}
&\Pr[(\nmExt(X_{<i}, Y, X_{>i}, \nmExt(X'_{<i}, f_i(Y), X'_{>i}))=(z, z')]-\Pr[(U_m, \nmExt(X'_{<i}, f_i(Y), X'_{>i}))=(z, z')] \\
=& \sum_{y \in B^+_{z, z'}} \Pr[Y=y] \Pr[(\nmExt(X_{<i}, y, X_{>i}, \nmExt(X'_{<i}, f_i(y), X'_{>i}))=(z, z')] \\ &-2^{-m}\sum_{y \in B^+_{z, z'}} \Pr[Y=y]\Pr[\nmExt(X'_{<i}, f_i(y), X'_{>i})=z']\\
=& \sum_{y \in B^+_{z, z'}} \Pr[Y=y] (\Pr[(\nmExt(X_{<i}, y, X_{>i}, \nmExt(X'_{<i}, f_i(y), X'_{>i}))=(z, z')] \\ &-2^{-m}\Pr[\nmExt(X'_{<i}, f_i(y), X'_{>i})=z']) \\
> & \e,
\end{align*}
which contradicts the fact that $\nmExt$ is a non-malleable extractor.

Now let $B=\cup_{z, z'}B_{z, z'}$, we have that $|B| \leq 2^{2m}2^{k+1}$. Thus, we now have that

\begin{align*}
&\left | (\nmExt(X_1, \cdots, X_s),  \nmExt(X'_1, \cdots, X'_s), X_i, X'_i)-(U_m, \nmExt(X'_1, \cdots, X'_s), X_i, X'_i) \right | \\
= & \sum_{y \in \zo^n}\Pr[X_i=y] \left |(\nmExt(X_{<i}, y, X_{>i}, \nmExt(X'_{<i}, f_i(y), X'_{>i}))-(U_m, \nmExt(X'_{<i}, f_i(y), X'_{>i})) \right |\\
\leq & \Pr[X_i \in B] \cdot 1+\Pr[X_i \notin B] 2^{2m} \e \\
\leq &  2^{2m} (\e+2^{k+1-k'})
\end{align*}
\end{proof}

We now have the following lemma.
\BL \label{lem:advcb2}
Suppose that there exists a constant $\gamma>0$ and an explicit construction of a strong non-malleable $s$-source extractor $\nmExt: (\zo^n)^s \to \zo^m$ for $(n, (1-2\gamma)n)$ sources which outputs $\Omega(n)$ bits with error $2^{-\Omega(n)}$. Then given any $t \in \N$ there is an explicit function $\acb: (\zo^n)^s \times \zo^{a} \to \zo^m$ with $m=\Omega(a)$ and the following property.

Let $X_1, \cdots, X_s$ be $s$ independent uniform strings on $n$ bits, and $\alpha, \alpha_1, \cdots, \alpha_t$ be $t+1$ strings on $a$ bits such that $\forall j \in [t], \alpha \neq \alpha_j$. Let $X^j_i, i \in [s], j \in [t]$ be random variables on $n$ bits such that $(\overline{X}_1=(X_1, \{X^j_1, j \in [t]\}), \cdots, \overline{X}_s=(X_s, \{X^j_s, j \in [t]\}))$ are independent (i.e., each $X^j_i$ only depends on $X_i$).  Let $Z=\acb(X_1, \cdots, X_s, \alpha)$ and $Z^j=\acb(X^j_1, \cdots, X^j_s, \alpha_j)$ for any $j \in [t]$. Then as long as $n \geq 2(t+1)^2 a/\gamma$, we have that $\forall i \in [s]$,

\[\left | (Z, \{Z^1, \cdots, Z^t\}, X_i)-(U_m, \{Z^1, \cdots, Z^t\}, X_i)\right | \leq ts2^{-\Omega(a)}.\]
\EL

We construct the function $\acb$ as follows. Let $\Ext$ be an optimal seeded extractor from Theorem~\ref{thm:optext} that uses $O(\log(n/\e))$ bits to extract from an $(n, k)$ source and output $0.9k$ bits. 
\begin{enumerate}
\item $\forall i \in [s]$, let $V_i$ be a slice of $X_i$ with length $a/\gamma$.
\item Repeat the following step for $t$ times: $\forall i \in [s]$, let $\tilde{V_i}=V_i \circ \alpha$. Compute $R=\nmExt(\tilde{V_1}, \cdots, \tilde{V_s})$. Then $\forall i \in [s]$, compute $V'_i=\Ext(X_i, R)$ and outputs $a/\gamma$ bits. Finally $\forall i \in [s]$, let $V_i=V'_i$.
\item Output $R$ from the last step, i.e., the computation of $V'_i=\Ext(X_i, R)$ and $V_i=V'_i$ in the above iteration can be omitted for the $t$'th execution.
\end{enumerate}

We now prove the lemma.
\begin{proof}
We prove the function $\acb$ described above is the desired function. We will use letters with superscript $j$ to denote random variables produced from $(X^j_i, i \in [s])$ and $\alpha_j$. By fixing additional randomness, without loss of generality we can assume that $\forall i \in [s]$, we have that $\forall j \in [t]$, $X^j_i$ is a deterministic function of $X_i$. We will use induction to prove the following claim.

\BCM
At the beginning of the $\ell$'th iteration, conditioned on the fixing of previous random variables (produced in previous rounds), we have that 
\begin{itemize}
\item $X_1, \cdots, X_s$ are still independent.
\item $\forall j \in [t]$, $V_i$ is a deterministic function of $X_i$ and $V^j_i$ is a deterministic function of $X^j_i$. 
\item $\forall i \in [s]$, the average conditional min-entropy of $X_i$ is at least $n-(\ell-1)(t+1)a/\gamma$.
\end{itemize}
At the end of the $\ell$'th iteration, we have that $\forall i$ and any $S \subseteq [t]$ with $|S| =\ell$, 

\[\left | (R, \{R^j, j \in S\}, V_i, \{V^j_i, j \in S\})-(U_m, \{R^j, j \in S\}, V_i, \{V^j_i, j \in S\})\right | \leq \ell s2^{-\Omega(a)}.\]
\ECM

To prove the claim, first note that since $\nmExt$ is a strong non-malleable $s$-source extractor $(n, (1-2\gamma)n)$ sources with error $2^{-\Omega(n)}$, it is also a strong non-malleable $s$-source extractor for average conditional min-entropy $(1-\gamma)n$ with error $2^{-\Omega(n)}$, by Lemma~\ref{entropies}. 

For the base case where $\ell=1$, clearly at the beginning of the first iteration, $X_1, \cdots, X_s$ are independent. Further, $\forall i \in s$, $V_i$ is a deterministic function of $X_i$ and $V^j_i$ is a deterministic function of $X^j_i$, $\forall j \in [t]$. Also the min-entropy of each $X_i$ is at least $n$. Now note that each $\tilde{V_i}$ has min-entropy rate at least $(a/\gamma)/(a+a/\gamma)> 1-\gamma$, and $\tilde{V_i} \neq \tilde{V}^j_i$ for any $j \in [t]$. Thus the claim follows by the assumption that $\nmExt$ is a strong non-malleable $s$-source extractor.

We next assume the claim holds for $\ell$ and show that it holds for $\ell+1$. The first three properties can be directly verified. We now prove the last property. Consider any set $S \subseteq [t]$ with $|S|=\ell+1$. Pick any $j_0 \in S$ and let $S' = S \setminus \{j_0\}$. By the claim we know that at the end of iteration $\ell$, we have that $\forall i \in [s]$, 

\[\left | (R, \{R^j, j \in S'\}, V_i, \{V^j_i, j \in S'\})-(U_m, \{R^j, j \in S'\}, V_i, \{V^j_i, j \in S'\})\right | \leq \ell s2^{-\Omega(a)}.\]

Consider any $i \in [s]$. We now fix $(V_i, \{V^j_i, j \in S'\})$. Note that all these random variables are deterministic functions of $(X_i, \{X^j_i, j \in S'\})$, which are in turn deterministic functions of $X_i$. Thus conditioned on this fixing, $X_1, \cdots, X_s$ are still independent. Also note that conditioned on this fixing, $(R, \{R^j, j \in S'\})$ is a deterministic function of $(\{V_h, h \neq i\}, \{V^j_h, h \neq i, j \in S'\})$, and therefore independent of $X_i$ and its derived random variables. Thus, we can further fix all the remaining $\{V^j_i, j \in [t]\}$ without affecting $(R, \{R^j, j \in S'\})$. Note that now the average conditional min-entropy of $X_i$ is at least $n-\ell(t+1)a/\gamma-(t+1) a/\gamma=n-(\ell+1)(t+1)a/\gamma$. 

Now we have that $R$ is still close to uniform given $\{R^j, j \in S'\}$. We now fix all $\{R^j, j \in S'\}$ and then all $\{V'^j_i=\Ext(X_i, R^j), j \in S'\}$. Note that fixing $\{R^j, j \in S'\}$ does not affect $X_i$, and conditioned on the fixing of  all $\{R^j, j \in S'\}$, we have that $\{V'^j_i=\Ext(X_i, R^j), j \in S'\}$ is a deterministic function of $(X_i, \{X^j_i, j \in S'\})$, which are in turn deterministic functions of $X_i$. Now the average conditional min-entropy of $X_i$ is at least $n-(\ell+1)(t+1)a/\gamma-\ell a/\gamma \geq n-(t+1)^2 a/\gamma>n/2$. Thus by Theorem~\ref{thm:optext} (and noticing that $R$ is still close to uniform and independent of $X_i$) we have 

\[|(V_i', R)-(U_m, R)| \leq 2^{-\Omega(a)}.\]

Note that given $R$, $V_i'$ is again a deterministic function of $X_i$. Thus (ignoring the $\ell s2^{-\Omega(a)}$ for now) we have the following inequality.

\[\left | (V_i', R, \{R^j, j \in S'\}, \{V'^j_i, j \in S'\}, \{V^j_i, j \in [t]\})-(U_m, R, \{R^j, j \in S'\}, \{V'^j_i, j \in S'\}, \{V^j_i, j \in [t]\}) \right | \leq 2^{-\Omega(a)}.\]

Furthermore, conditioned on the fixing of $(R, \{R^j, j \in S'\}, \{V'^j_i, j \in S'\}, \{V^j_i, j \in [t]\})$, $V_i'$ is a deterministic function of $X_i$ and therefore independent of $\{X_h, h \neq i\}$. Thus, we can also fix all the other $\{V^j_h, j \in [t], h \neq i\}$ without affecting the inequality. Thus we obtain the following.

\[\left | (V_i', \{V'^j_i, j \in S'\}, \{V^j_h, j \in [t], h \in [s]\})-(U_m, \{V'^j_i, j \in S'\}, \{V^j_h, j \in [t], h \in [s]\}) \right | \leq 2^{-\Omega(a)}.\]

Using the same argument, we can also show that conditioned on the fixing of $(\{V'^j_i, j \in S'\}, \{V^j_h, j \in [t], h \in [s]\})$, $V'^{j_0}_i$ is a deterministic function of $X^{j_0}_i$, which in turn is a deterministic function of $X_i$. However, we don't know if $V'^{j_0}_i$ is close to uniform, and it may be correlated with $V_i'$.

We can repeat the above argument for any $i \in [s]$, thus we obtain the following conclusion.

\begin{itemize}
\item $\forall i \in [s]$, we have
\[\left | (V_i', \{V'^j_i, j \in S'\}, \{V^j_h, j \in [t], h \in [s]\})-(U_m, \{V'^j_i, j \in S'\}, \{V^j_h, j \in [t], h \in [s]\}) \right | \leq 2^{-\Omega(a)}.\]
\item Further, $\forall i \in [s]$, conditioned on the fixing of $(\{V'^j_i, j \in S'\}, \{V^j_h, j \in [t], h \in [s]\})$, we have that $(V_i', V'^{j_0}_i)$ is a deterministic function of $X_i$.
\end{itemize}

Now fix $(\{V'^j_i, j \in S'\}, \{V^j_h, j \in [t], h \in [s]\})$. Note that conditioned on this fixing,  $X_1, \cdots, X_s$ are still independent. Thus $(V_i', V'^{j_0}_i)$ are also independent.
By the fact that $\nmExt$ is a strong non-malleable $s$-source extractor, we have that $\forall i \in [s]$,

\[\left | (R, R^{j_0}, V_i', V'^{j_0}_i)-(U_m, R^{j_0}, V_i', V'^{j_0}_i)\right | \leq 2^{-\Omega(a)}.\]

Since we have fixed all the $(\{V'^j_i, j \in S'\})$ before, and each new $(R^j, j \in S')$ is now a deterministic function of $(\{V'^j_i, j \in S'\})$, by adding back all the errors we also have that

\[\left | (R, \{R^j, j \in S\}, V_i', \{V'^j_i, j \in S\})-(U_m, \{R^j, j \in S\}, V_i', \{V'^j_i, j \in S\})\right | \leq (\ell+1)s 2^{-\Omega(a)}.\]
Note that at the end of iteration we replace $V_i$ with $V'_i$, so the claim holds and the theorem is proved.
\end{proof}

We now have the following theorem.

\BT \label{thm:nmconvert1}
Suppose there is a constant $\gamma>0$ and an explicit non-malleable $(s+1)$-source extractor for $(n, (1-\gamma)n)$ sources with error $2^{-\Omega(n)}$ and output length $\Omega(n)$. Then there is a constant $C>0$ such that for any $0< \e < 1$ with $k \geq C t^2\log(n/\e)$, there is an explicit strong seeded $t$-non-malleable extractor for $s$ independent $(n, k)$ sources with seed length $d=C t^2\log(n/\e)$, error $O(t s\e)$ and output length $\Omega(\log(1/\e))$. 
\ET

The construction of the seeded non-malleable extractor for $s$ independent $(n, k)$ sources is as follows. Let the sources be $X_1, \cdots, X_s$ and the seed be $Y$.

\begin{itemize}
\item Let $\Ext: \zo^n \times \zo^{d'} \to \zo^m$ be an optimal seeded extractor from Theorem~\ref{thm:optext}, which uses $d_1=O(\log(n/\e))$ random bits to extract from $(n, k)$ sources and output $m=0.9k$ bits.

\item Let $\Ext_1, \Ext_2$ be optimal seeded extractors from Theorem~\ref{thm:optext}.

\item Let $\bip$ be the inner product two-source extractor from Theorem~\ref{thm:ip}.

\item Let $\acb$ be the correlation breaker with advice from Lemma~\ref{lem:advcb2}.

\item Let $\adg$ be the advice generator from Theorem~\ref{adv_gen}.
\end{itemize}

\begin{enumerate}
\item Take a small slice $Y_1$ of $Y$ with length $d_1=O(\log(n/\e))$, for every $i \in [s]$, compute $Z_i=\Ext'(X_i, Y_1)$ which outputs $0.9k$ bits. 

\item Use $Z_1$ and $X$ to compute $\adg(X, Y)$. Specifically, as in Theorem~\ref{adv_gen}, take a small slice $\overline{Z}_1$ of $Z_1$ with length $d_2=O(\log(n/\e))$ and compute $Y_2=g(Y, \overline{Z}_1)$ which outputs $d_3=O(\log(1/\e))$ bits. Let $\adg(X_1, Y)=(Y_1, Y_2)=\alpha$. Note that the length of the advice is $a=d_1+d_3=O(\log(n/\e))$. We choose the hidden constant to be big enough such that the term $2^{-\Omega(a)}$ in Lemma~\ref{lem:advcb2} is at most $\e$.

\item Let $d_4=max\{d_2, d_1+d_3\}$. Take a slice of $Y_3$ of $Y$ with length $d_5=3(t+1)d_4$, and a slice $Z_3$ of $Z$ with length $d_5=3td_4$. Compute $R=\bip(Y_3, Z_3)$. 

\item Compute $\tilde{Y}=\Ext_1(Y, R)$ which outputs $m_1=0.5d$ bits, and $\tilde{Z_1}=\Ext_2(Z_1, R)$ which outputs $m_1$ bits. For $i=2, \cdots, s$, truncate each $Z_i$ to $\tilde{Z_i}$ with $m_1$ bits. 

\item Output $V=\acb(\tilde{Y}, \tilde{Z_1}, \cdots, \tilde{Z_s}, \alpha)$.
\end{enumerate}

\begin{proof}
Suppose we have $t$ tampered seeds $Y^j =f_j(Y), j \in [t]$, where each $f_j$ has no fixed points. We will use letters with superscript $j$ to denote random variables obtained from $(Y^j, X_1, \cdots, X_s)$. First, by Theorem~\ref{thm:optext}, we have that for any $i \in [s]$, 
\[|(Z_i, Y_1)-(U, Y_1)| \leq \e.\]

Since conditioned on the fixing of $Y_1$, each $Z_i$ is a deterministic function of $X_i$ and thus independent, we have

\[|(Z_1, \cdots, Z_s, Y_1)-(U, \cdots, U, Y_1)| \leq s\e.\]

We will now proceed as if $(Z_1, \cdots, Z_s)$ are uniform and independent, given $Y_1$. Take any $j \in [t]$, by Theorem~\ref{adv_gen}, we know that with probability $1-\e$ over the fixing of $(Y_1, \overline{Z}_1, Y_2, Y^j_1, \overline{Z}^j_1, Y^j_2)$, we have $\alpha \neq \alpha^j$. Thus, with probability $1-t \e$ over the fixing of $H=(Y_1, \overline{Z}_1, Y_2, \{Y^j_1, \overline{Z}^j_1, Y^j_2, j \in [t]\})$, we have that $\forall j, \alpha \neq \alpha^j$. Furthermore, notice that conditioned on the fixing of $H$, we have that $(Y, \{Y^j, j \in [t]\}$ and $(Z_1, \{Z^j_1, j \in [t]\})$ are still independent, the average conditional min-entropy of $Y_3$ is at least $d_5-(t+1)(d_1+d_3) \geq 2(t+1)d_4$, and the average conditional min-entropy of $Z_3$ is at least $d_5-(t+1)d_2 \geq 2(t+1)d_4$. Also note that the fixing of $H$ does not affect $Z_2, \{Z^j_2, j \in [t]\}, \cdots, Z_s, \{Z^j_s, j \in [t]\}$.

Now by Theorem~\ref{thm:ip}, we have that 

\[|(R, Y_3)-(U, Y_3)| \leq \e.\]

Note that conditioned on the fixing of $(Y_3, \{Y^j_3, j \in [t]\})$, $(R, \{R^j, j \in [t]\})$ is a deterministic function of $(Z_3, \{Z^j_3, j \in [t]\})$, and thus independent of $(Y, \{Y^j, j \in [t]\})$. Moreover $R$ is close to uniform and the average conditional min-entropy of $Y$ is at least $d-(t+1)(d_1+d_3+d_5)=d-O(t^2 \log(n/\e))$. Thus by taking $C$ to be large enough we have that $d-O(t^2 \log(n/\e)) > 2d/3$. Thus by Theorem~\ref{thm:optext} we have that 

\[|(\tilde{Y}, R)-(U, R)| \leq \e.\]

Note that conditioned on the further fixing of $(R, \{R^j, j \in [t]\})$, $(\tilde{Y}, \{\tilde{Y}^j, j \in [t]\})$ is a deterministic function of $(Y, \{Y^j, j \in [t]\})$. Thus we can further fix $(Z_3, \{Z^j_3, j \in [t]\})$ without affecting the above inequality. Similarly, we also have 

\[|(R, Z_3)-(U, Z_3)| \leq \e.\]

Note that conditioned on the fixing of $(Z_3, \{Z^j_3, j \in [t]\})$, $(R, \{R^j, j \in [t]\})$ is a deterministic function of $(Y_3, \{Y^j_3, j \in [t]\}$, and thus independent of $(Z_1, \{Z^j_1, j \in [t]\}$. Moreover $R$ is close to uniform and the average conditional min-entropy of $Z_1$ is at least $0.9k-(t+1)(d_2+d_5)=0.9k-O(t^2 \log(n/\e))$. Thus by taking $C$ to be large enough we have that $0.9k-O(t^2 \log(n/\e)) > 2d/3$. Thus by Theorem~\ref{thm:optext} we have that 

\[|(\tilde{Z}_1, R)-(U, R)| \leq \e.\]

Note that conditioned on the further fixing of $(R, \{R^j, j \in [t]\})$, $(\tilde{Z}_1, \{\tilde{Z}^j_1, j \in [t]\})$ is a deterministic function of $(Z_1, \{Z^j_1, j \in [t]\})$. Thus we can further fix $(Y_3, \{Y^j_3, j \in [t]\})$ without affecting the above inequality. Note that none of these affects $Z_2, \{Z^j_2, j \in [t]\}, \cdots, Z_s, \{Z^j_s, j \in [t]\}$. Therefore, combining the above we have that with probability $1-O(t \e)$ over the fixing of $\bar{H}=(H, Z_3, \{Z^j_3, j \in [t]\}, R, \{R^j, j \in [t]\}, Y_3, \{Y^j_3, j \in [t]\})$,
\begin{itemize}
\item $\forall j, \alpha \neq \alpha^j$.
\item $(\tilde{Y}, \{\tilde{Y}^j, j \in [t]\}), (\tilde{Z_1}, \{\tilde{Z}^j_1, j \in [t]\}), \cdots, (\tilde{Z_s}, \{\tilde{Z}^j_s, j \in [t]\})$ are independent.
\item $(\tilde{Y}, \tilde{Z_1}, \cdots, \tilde{Z_s}) \approx_{O(s \e)} (U_{m_1}, \cdots, U_{m_1}).$
\end{itemize}

Next, using Theorem~\ref{thm:nmstrong}, we see that the non-malleable $(s+1)$-source extractor is also a strong non-malleable $(s+1)$-source extractor for $(n, (1-\gamma/2)n)$ sources with error $2^{2m'}(2^{-\Omega(n)}+2^{1-\gamma n/2})$, where $m'$ is the output length of the extractor. By truncating the output if necessary, we can ensure that $m'=\Omega(n)$ and $2^{2m'}(2^{-\Omega(n)}+2^{1-\gamma n/2})=2^{-\Omega(n)}$. Thus the non-malleable $(s+1)$-source extractor is also a strong non-malleable $(s+1)$-source extractor for $(n, (1-\gamma/2)n)$ sources with error $2^{-\Omega(n)}$ and output length $\Omega(n)$.

We now apply Lemma~\ref{lem:advcb2}. First ignoring the error, and note that the length of each $(\tilde{Y}, \tilde{Z_1}, \cdots, \tilde{Z_s})$ is $0.5d$ where $d=C t^2\log(n/\e)$, and the length of advice is $a=O(\log(n/\e))$. By choosing the constant $C$ large enough we can ensure that $0.5d \geq 2(t+1)^2 a/(\gamma/4)$. Therefore by Lemma~\ref{lem:advcb2}, we have that the output has length $\Omega(\log(1/\e))$, and $\forall i$, 

\[\left | (V, \{V^1, \cdots, V^t\}, X_i)-(U_m, \{V^1, \cdots, V^t\}, X_i)\right | \leq t(s+1)\e.\]

Adding back all the errors we see that the construction is a seeded $t$-non-malleable extractor for $s$ independent $(n, k)$ sources with error $O(t s\e)$ and output length $\Omega(\log(1/\e))$.
\end{proof}

The above construction and theorem can also be easily generalized to the case where we do not have non-malleable $s+1$-source extractors with asymptotically optimal error. For example, suppose to get error $\e$ the non-malleable $s+1$-source extractor needs $(f(\e), (1-\gamma)f(\e))$ sources for some function $f$ (note that $f(\e)$ is at least $O(\log(1/\e))$, then all we need to change is that in Lemma~\ref{lem:advcb2}, the size of each $V_i$ should become $O(\log n+f(\e))$. Since the length of the advice is always going to be $O(\log(n/\e))$, this ensures that each time when we apply the non-malleable $(s+1)$-source extractor, the sources have entropy rate at least $1-\gamma$ and error $\e$. Now the same analysis in Theorem~\ref{thm:nmconvert1} goes through, as long as $k, d \geq C t^2(\log n +f(\e))$. Thus, we have the following theorem.

\BT \label{thm:nmconvert2}
Suppose there is a function $f$, a constant $\gamma>0$ and an explicit non-malleable $(s+1)$-source extractor for $(f(\e), (1-\gamma)f(\e))$ sources with error $\e$ and output length $\Omega(f(\e))$. Then there is a constant $C>0$ such that for any $0< \e < 1$ with $k \geq C t^2(\log n +f(\e))$, there is an explicit strong seeded $t$-non-malleable extractor for $s$ independent $(n, k)$ sources with seed length $d=Ct^2(\log n +f(\e))$, error $O(t s\e)$ and output length $\Omega(f(\e))$. 
\ET

The constructions and theorems can also be extended to the case of $t$-non-malleable extractors for $s$ independent sources, we omit the details for now.

We now combine Theorem~\ref{thm:nmconvert1} and Theorem~\ref{thm:nmconvert2} with known constructions of non-malleable $s$-source extractors to obtain seeded $t$-non-malleable extractors. By combining Theorem~\ref{thm:nmconvert2} and Theorem~\ref{thm:nmtext}, we have the following theorem (note that here $f(\e)=O(\log(1/\e)\log \log (1/\e))$.

\BT \label{thm:tnm1}
There exists a constant $C>1$ such that for any $t \in \N$, $0<\e<1$ and $k \geq C t^2(\log n +\log(1/\e)\log \log (1/\e))$, there is an explicit strong seeded $t$-non-malleable extractor for $(n, k)$ sources with seed length $d=C t^2(\log n +\log(1/\e)\log \log (1/\e))$, output length $\Omega(\log(1/\e)\log \log (1/\e))$ and error $O(t \e)$.
\ET

Next, we use the following theorem proved by Chattopadhyay and Zuckerman \cite{CZ14}.

\BT [\cite{CZ14}]
There is a constant $0<\gamma<1$ and an explicit non-malleable $10$-source extractor for $(n, (1-\gamma)n)$ sources with error $2^{-\Omega(n)}$ and output length $\Omega(n)$.
\ET

Combining this theorem with Theorem~\ref{thm:nmconvert1}, we have the following theorem.

\BT \label{thm:tnm2}
There exists a constant $C>1$ such that for any $t \in \N$, $0<\e<1$ and $k \geq C t^2(\log (n/\e))$, there is an explicit strong seeded $t$-non-malleable extractor for $9$ independent $(n, k)$ sources with seed length $d=C t^2(\log (n/\e))$, output length $\Omega(\log(1/\e))$ and error $O(t \e)$.
\ET

By using improved somewhere random condensers as samplers and following the framework in \cite{CZ15}, Ben-Aroya et. al \cite{BDT16} proved the following theorem.

\BT \cite{BDT16} \label{thm:gext}
Suppose there is a function $f$ and an explicit strong seeded $t$-non-malleable extractor for $s$ independent $(n, k')$ sources with seed length and entropy requirement $d=k'=f(t, \e)$, then there for every constant $\e>0$ exist constants $t=t(\e), c=c(\e)$ and an explicit extractor $\Ext: (\zo^n)^s \to \zo$ for $s$ independent $(n, k)$ sources with $k \geq f(t, 1/n^c)$ and error $\e$.
\ET

\begin{remark}
The original construction in \cite{BDT16} is just for two sources, but it extends directly to any $s$ sources just by treating $s-1$ sources as one source. 
\end{remark}

We can now use above theorems to get improved constructions of independent source extractors. For example, combining the above theorem with Theorem~\ref{thm:tnm1}, we immediately obtain the following theorem.

\BT \label{thm:2source}
For every constant $\e>0$ exists a constant $c>1$ and an explicit two-source extractor $\Ext: (\zo^n)^2 \to \zo$ for min-entropy $k \geq c \log n \log \log n$, with error $\e$.
\ET

Using Theorem~\ref{thm:tnm2} instead, we obtain the following theorem.

\BT \label{thm:10source}
For every constant $\e>0$ exists a constant $c>1$ and an explicit ten-source extractor $\Ext: (\zo^n)^{10} \to \zo$ for min-entropy $k \geq c \log n$, with error $\e$.
\ET

%% file: optext.bbl
\newcommand{\etalchar}[1]{$^{#1}$}
\begin{thebibliography}{BADTS16}

\bibitem[ADKO15]{ADKO15}
D.~Aggarwal, Y.~Dodis, T.~Kazana, and M.~Obremski.
\newblock Non-malleable reductions and applications.
\newblock In {\em Proceedings of the 47th Annual ACM Symposium on Theory of
  Computing}, 2015.

\bibitem[ADL14]{ADL14}
Divesh Aggarwal, Yevgeniy Dodis, and Shachar Lovett.
\newblock Non-malleable codes from additive combinatorics.
\newblock In {\em Proceedings of the 46th Annual ACM Symposium on Theory of
  Computing}, 2014.

\bibitem[Agg14]{Agw14}
Divesh Aggarwal.
\newblock Affine-evasive sets modulo a prime.
\newblock Technical Report 2014/328, Cryptology ePrint Archive, 2014.

\bibitem[BADTS16]{BDT16}
Avraham Ben-Aroya, Dean Doron, and Amnon Ta-Shma.
\newblock Explicit two-source extractors for near-logarithmic min-entropy.
\newblock Technical Report TR16-088, ECCC, 2016.

\bibitem[BBR88]{BennettBR88}
Charles~H. Bennett, Gilles Brassard, and {Jean-Marc} Robert.
\newblock Privacy amplification by public discussion.
\newblock {\em SIAM Journal on Computing}, 17(2):210--229, April 1988.

\bibitem[BIW04]{BarakIW04}
Boaz Barak, R.~Impagliazzo, and Avi Wigderson.
\newblock Extracting randomness using few independent sources.
\newblock In {\em Proceedings of the 45th Annual IEEE Symposium on Foundations
  of Computer Science}, pages 384--393, 2004.

\bibitem[BKS{\etalchar{+}}05]{BarakKSSW05}
Boaz Barak, Guy Kindler, Ronen Shaltiel, Benny Sudakov, and Avi Wigderson.
\newblock Simulating independence: New constructions of condensers, {R}amsey
  graphs, dispersers, and extractors.
\newblock In {\em Proceedings of the 37th Annual ACM Symposium on Theory of
  Computing}, pages 1--10, 2005.

\bibitem[Bou05]{Bourgain05}
Jean Bourgain.
\newblock More on the sum-product phenomenon in prime fields and its
  applications.
\newblock {\em International Journal of Number Theory}, 1:1--32, 2005.

\bibitem[BRSW06]{BarakRSW06}
Boaz Barak, Anup Rao, Ronen Shaltiel, and Avi Wigderson.
\newblock 2 source dispersers for {$n^{o(1)}$} entropy and {R}amsey graphs
  beating the {F}rankl-{W}ilson construction.
\newblock In {\em Proceedings of the 38th Annual ACM Symposium on Theory of
  Computing}, 2006.

\bibitem[CG88]{ChorG88}
Benny Chor and Oded Goldreich.
\newblock Unbiased bits from sources of weak randomness and probabilistic
  communication complexity.
\newblock {\em SIAM Journal on Computing}, 17(2):230--261, 1988.

\bibitem[CG14a]{CG14a}
Mahdi Cheraghchi and Venkatesan Guruswami.
\newblock Capacity of non-malleable codes.
\newblock In {\em ITCS}, pages 155--168, 2014.

\bibitem[CG14b]{CG14b}
Mahdi Cheraghchi and Venkatesan Guruswami.
\newblock Non-malleable coding against bit-wise and split-state tampering.
\newblock In {\em TCC}, pages 440--464, 2014.

\bibitem[CGL16]{CGL15}
Eshan Chattopadhyay, Vipul Goyal, and Xin Li.
\newblock Non-malleable extractors and codes, with their many tampered
  extensions.
\newblock In {\em Proceedings of the 48th Annual ACM Symposium on Theory of
  Computing}, 2016.

\bibitem[CKOR10]{ckor}
N.~Chandran, B.~Kanukurthi, R.~Ostrovsky, and L.~Reyzin.
\newblock Privacy amplification with asymptotically optimal entropy loss.
\newblock In {\em Proceedings of the 42nd Annual ACM Symposium on Theory of
  Computing}, pages 785--794, 2010.

\bibitem[CL16]{CL16}
Eshan Chattopadhyay and Xin Li.
\newblock Explicit non-malleable extractors, multi-source extractors and almost
  optimal privacy amplification protocols.
\newblock In {\em Proceedings of the 57th Annual IEEE Symposium on Foundations
  of Computer Science}, 2016.

\bibitem[Coh15]{Cohen15}
Gil Cohen.
\newblock Local correlation breakers and applications to three-source
  extractors and mergers.
\newblock In {\em Proceedings of the 56th Annual IEEE Symposium on Foundations
  of Computer Science}, 2015.

\bibitem[Coh16a]{Coh16}
Gil Cohen.
\newblock Making the most of advice: New correlation breakers and their
  applications.
\newblock In {\em Proceedings of the 57th Annual IEEE Symposium on Foundations
  of Computer Science}, 2016.

\bibitem[Coh16b]{Coh15nm}
Gil Cohen.
\newblock Non-malleable extractors - new tools and improved constructions.
\newblock In {\em Proceedings of the 31st Annual IEEE Conference on
  Computational Complexity}, 2016.

\bibitem[Coh16c]{Coh16a}
Gil Cohen.
\newblock Non-malleable extractors with logarithmic seeds.
\newblock Technical Report TR16-030, ECCC, 2016.

\bibitem[Coh16d]{Cohen16}
Gil Cohen.
\newblock Two-source extractors for quasi-logarithmic min-entropy and improved
  privacy amplification protocols.
\newblock Technical Report TR16-114, ECCC: Electronic Colloquium on
  Computational Complexity, 2016.

\bibitem[CRS14]{CRS11}
Gil Cohen, Ran Raz, and Gil Segev.
\newblock Non-malleable extractors with short seeds and applications to privacy
  amplification.
\newblock {\em SIAM Journal on Computing}, 43(2):450--476, 2014.

\bibitem[CS16]{Coh16b}
Gil Cohen and Leonard Schulman.
\newblock Extractors for near logarithmic min-entropy.
\newblock In {\em Proceedings of the 57th Annual IEEE Symposium on Foundations
  of Computer Science}, 2016.

\bibitem[CZ14]{CZ14}
Eshan Chattopadhyay and David Zuckerman.
\newblock Non-malleable codes against constant split-state tampering.
\newblock In {\em Proceedings of the 55th Annual IEEE Symposium on Foundations
  of Computer Science}, pages 306--315, 2014.

\bibitem[CZ16]{CZ15}
Eshan Chattopadhyay and David Zuckerman.
\newblock Explicit two-source extractors and resilient functions.
\newblock In {\em Proceedings of the 48th Annual ACM Symposium on Theory of
  Computing}, 2016.

\bibitem[DKO13]{DKO13}
Stefan Dziembowski, Tomasz Kazana, and Maciej Obremski.
\newblock Non-malleable codes from two-source extractors.
\newblock In {\em CRYPTO (2)}, pages 239--257, 2013.

\bibitem[DKRS06]{dkrs}
Y.~Dodis, J.~Katz, L.~Reyzin, and A.~Smith.
\newblock Robust fuzzy extractors and authenticated key agreement from close
  secrets.
\newblock In {\em Advances in Cryptology --- CRYPTO '06, 26th Annual
  International Cryptology Conference, Proceedings}, pages 232--250, 2006.

\bibitem[DKSS09]{DvirKSS09}
Zeev Dvir, Swastik Kopparty, Shubhangi Saraf, and Madhu Sudan.
\newblock Extensions to the method of multiplicities, with applications to
  kakeya sets and mergers.
\newblock In {\em Proceedings of the 50th Annual IEEE Symposium on Foundations
  of Computer Science}, 2009.

\bibitem[DLWZ14]{DLWZ11}
Yevgeniy Dodis, Xin Li, Trevor~D. Wooley, and David Zuckerman.
\newblock Privacy amplification and non-malleable extractors via character
  sums.
\newblock {\em SIAM Journal on Computing}, 43(2):800--830, 2014.

\bibitem[DORS08]{dors}
Y.~Dodis, R.~Ostrovsky, L.~Reyzin, and A.~Smith.
\newblock Fuzzy extractors: How to generate strong keys from biometrics and
  other noisy data.
\newblock {\em SIAM Journal on Computing}, 38:97--139, 2008.

\bibitem[DP07]{DP07}
Stefan Dziembowski and Krzysztof Pietrzak.
\newblock Intrusion-resilient secret sharing.
\newblock In {\em Proceedings of the 48th Annual IEEE Symposium on Foundations
  of Computer Science}, FOCS '07, pages 227--237, Washington, DC, USA, 2007.
  IEEE Computer Society.

\bibitem[DPW10]{DPW10}
Stefan Dziembowski, Krzysztof Pietrzak, and Daniel Wichs.
\newblock Non-malleable codes.
\newblock In {\em ICS}, pages 434--452, 2010.

\bibitem[DW08]{DvirW08}
Zeev Dvir and Avi Wigderson.
\newblock Kakeya sets, new mergers and old extractors.
\newblock In {\em Proceedings of the 49th Annual IEEE Symposium on Foundations
  of Computer Science}, 2008.

\bibitem[DW09]{DW09}
Yevgeniy Dodis and Daniel Wichs.
\newblock Non-malleable extractors and symmetric key cryptography from weak
  secrets.
\newblock In {\em Proceedings of the 41st Annual ACM Symposium on Theory of
  Computing}, pages 601--610, 2009.

\bibitem[DY13]{DY12}
Yevgeniy Dodis and Yu~Yu.
\newblock Overcoming weak expectations.
\newblock In {\em 10th Theory of Cryptography Conference}, 2013.

\bibitem[GUV09]{GuruswamiUV09}
Venkatesan Guruswami, Christopher Umans, and Salil Vadhan.
\newblock Unbalanced expanders and randomness extractors from
  {P}arvaresh-{V}ardy codes.
\newblock {\em Journal of the ACM}, 56(4), 2009.

\bibitem[KR09]{KR09}
B.~Kanukurthi and L.~Reyzin.
\newblock Key agreement from close secrets over unsecured channels.
\newblock In {\em EUROCRYPT 2009, 28th Annual International Conference on the
  Theory and Applications of Cryptographic Techniques}, 2009.

\bibitem[Li11]{Li11b}
Xin Li.
\newblock Improved constructions of three source extractors.
\newblock In {\em Proceedings of the 26th Annual IEEE Conference on
  Computational Complexity}, pages 126--136, 2011.

\bibitem[Li12a]{Li12a}
Xin Li.
\newblock Design extractors, non-malleable condensers and privacy
  amplification.
\newblock In {\em Proceedings of the 44th Annual ACM Symposium on Theory of
  Computing}, pages 837--854, 2012.

\bibitem[Li12b]{Li12b}
Xin Li.
\newblock Non-malleable extractors, two-source extractors and privacy
  amplification.
\newblock In {\em Proceedings of the 53rd Annual IEEE Symposium on Foundations
  of Computer Science}, pages 688--697, 2012.

\bibitem[Li13a]{Li13b}
Xin Li.
\newblock Extractors for a constant number of independent sources with
  polylogarithmic min-entropy.
\newblock In {\em Proceedings of the 54th Annual IEEE Symposium on Foundations
  of Computer Science}, pages 100--109, 2013.

\bibitem[Li13b]{Li13a}
Xin Li.
\newblock New independent source extractors with exponential improvement.
\newblock In {\em Proceedings of the 45th Annual ACM Symposium on Theory of
  Computing}, pages 783--792, 2013.

\bibitem[Li15a]{Li15b}
Xin Li.
\newblock Non-malleable condensers for arbitrary min-entropy, and almost
  optimal protocols for privacy amplification.
\newblock In {\em 12th IACR Theory of Cryptography Conference}, pages 502--531.
  Springer-Verlag, 2015.
\newblock LNCS 9014.

\bibitem[Li15b]{Li15}
Xin Li.
\newblock Three source extractors for polylogarithmic min-entropy.
\newblock In {\em Proceedings of the 56th Annual IEEE Symposium on Foundations
  of Computer Science}, 2015.

\bibitem[Li16]{Li16}
Xin Li.
\newblock Improved two-source extractors, and affine extractors for
  polylogarithmic entropy.
\newblock In {\em Proceedings of the 57th Annual IEEE Symposium on Foundations
  of Computer Science}, 2016.

\bibitem[LRVW03]{LuRVW03}
C.~J. Lu, Omer Reingold, Salil Vadhan, and Avi Wigderson.
\newblock Extractors: Optimal up to constant factors.
\newblock In {\em Proceedings of the 35th Annual ACM Symposium on Theory of
  Computing}, pages 602--611, 2003.

\bibitem[Mek15]{Mek:resil}
Raghu Meka.
\newblock Explicit resilient functions matching {Ajtai-Linial}.
\newblock {\em CoRR}, abs/1509.00092, 2015.

\bibitem[MW97]{MW97}
Ueli~M. Maurer and Stefan Wolf.
\newblock Privacy amplification secure against active adversaries.
\newblock In {\em Advances in Cryptology --- CRYPTO '97, 17th Annual
  International Cryptology Conference, Proceedings}, 1997.

\bibitem[NZ96]{NisanZ96}
Noam Nisan and David Zuckerman.
\newblock Randomness is linear in space.
\newblock {\em Journal of Computer and System Sciences}, 52(1):43--52, 1996.

\bibitem[Rao06]{Rao06}
Anup Rao.
\newblock Extractors for a constant number of polynomially small min-entropy
  independent sources.
\newblock In {\em Proceedings of the 38th Annual ACM Symposium on Theory of
  Computing}, 2006.

\bibitem[Raz05]{Raz05}
Ran Raz.
\newblock Extractors with weak random seeds.
\newblock In {\em Proceedings of the 37th Annual ACM Symposium on Theory of
  Computing}, pages 11--20, 2005.

\bibitem[RW03]{RW03}
Renato Renner and Stefan Wolf.
\newblock Unconditional authenticity and privacy from an arbitrarily weak
  secret.
\newblock In {\em Advances in Cryptology --- CRYPTO '03, 23rd Annual
  International Cryptology Conference, Proceedings}, pages 78--95, 2003.

\bibitem[Vad04]{Vadhan04}
Salil~P. Vadhan.
\newblock Constructing locally computable extractors and cryptosystems in the
  bounded-storage model.
\newblock {\em J. Cryptology}, 17(1):43--77, 2004.

\end{thebibliography}
